\documentclass[twocolumn]{aastex63}

\usepackage{apjfonts}
\usepackage{natbib}
\usepackage{amssymb}
\usepackage{enumitem}
\usepackage{amsmath}

\usepackage{iondefs}
\defcitealias{kacprzak15}{KMC15}

\usepackage[flushleft]{threeparttable}

\providecommand{\e}[1]{\ensuremath{\times 10^{#1}}}

\hypersetup{breaklinks,colorlinks,citecolor=blue,linkcolor=magenta}

\shorttitle{Distribution of {\OVI} Covering Fractions}
\shortauthors{\sc Marra {\etal}}

\begin{document}

\title{\sc Spatial Distribution of {\OVI} Covering Fractions in the Simulated Circumgalactic Medium}

\author[0000-0002-8362-0517]{Rachel Marra}
\affiliation{Department of Astronomy, New Mexico State University, Las Cruces, NM 88003, USA\\}

\author[0000-0002-8362-0517]{Christopher W. Churchill}
\affiliation{Department of Astronomy, New Mexico State University, Las Cruces, NM 88003, USA\\}

\author[0000-0003-1362-9302]{Glenn G. Kacprzak}
\affiliation{Centre for Astrophysics and Supercomputing, Swinburne University of Technology, Hawthorn, Victoria 3122, Australia\\}
\affiliation{ARC Centre of Excellence for All Sky Astrophysics in 3  Dimensions (ASTRO 3D)}

\author{Rachel Vander Vliet}
\affiliation{Sofia Science Center at NASA Ames, Moffett Field, California}

\author[0000-0002-8680-248X]{Daniel Ceverino}
\affiliation{Universidad Autónoma de Madrid, Ciudad Universitaria de Cantoblanco, 28049 Madrid, Spain}
\affiliation{CIAFF, Facultad de Ciencias, Universidad Autonoma de Madrid, 28049 Madrid, Spain}

\author[0000-0003-0091-952X]{Emmy Lewis}
\affiliation{Department of Mathematics, New Mexico State University, Las Cruces, NM 88003, USA\\}

\author[0000-0003-2377-8352]{Nikole M. Nielsen}
\affiliation{Centre for Astrophysics and Supercomputing, Swinburne University of Technology, Hawthorn, Victoria 3122, Australia\\}
\affiliation{ARC Centre of Excellence for All Sky Astrophysics in 3  Dimensions (ASTRO 3D)}

\author[0000-0003-3938-8762]{Sowgat Muzahid}
\affiliation{Leibniz-Institute for Astrophysics Potsdam (AIP), An der Sternwarte 16, D-14482 Potsdam, Germany}

\author[0000-0003-4877-9116]{Jane C. Charlton}
\affiliation{Department of Astronomy and Astrophysics, The Pennsylvania State University, State College, PA 16801, USA}

\begin{abstract}
We use adaptive mesh refinement cosmological simulations to study the spatial distribution and covering fraction of {\OVI} absorption in the circumgalactic medium (CGM) as a function of projected virial radius and azimuthal angle.  We compare these simulations to an observed sample of 53 galaxies from the Multiphase Galaxy Halos Survey. Using {\sc Mockspec}, an absorption line analysis pipeline, we generate synthetic quasar absorption line observations of the simulated CGM. To best emulate observations, we studied the averaged properties of 15,000 ``mock samples'' each of 53 sightlines having a distribution of $D/R_{vir}$ and sightline orientation statistically consistent with the observations. We find that the {\OVI} covering fraction obtained for the simulated galaxies agrees well with the observed value for the inner halo ($D/R_{vir} \leq 0.375$) and is within $1.1\sigma$ in the outer halo ($D/R_{vir} > 0.75$), but is underproduced within $0.375 < D/R_{vir} \leq 0.75$. The observed bimodal distribution of {\OVI} covering fraction with azimuthal angle, showing higher frequency of absorption along the projected major and minor axes of galaxies, is not reproduced in the simulations. Further analysis reveals the spatial-kinematic distribution of {\OVI} absorbing gas is dominated by outflows in the inner halo mixed with a inflowing gas that originates from further out in the halo.  Though the CGM of the individual simulated galaxies exhibit spatial structure, the flat azimuthal distribution occurs because the individual simulated galaxies do not develop a CGM structure that is universal from galaxy to galaxy.
 \end{abstract}
\keywords{galaxies: quasars: absorption lines}

\section{Introduction}

Throughout the cosmic web, filaments of dark matter intersect to form dark matter halos \citep[e.g.,][]{Bond96, VanBond08}. Baryons gravitationally follow the dark matter overdensities and, if there is a high enough density of baryons, they will collapse and start forming stars. Thus, it is in these dark matter halos that galaxies form \citep[e.g.][]{Doroshkevich80, Klypin83, Shapiro83, white91, Pauls95, Sathyaprakash96}.

Depending on the local cosmic environment and the mass of the dark matter halo, different galaxy types (i.e., dwarf, elliptical, spiral) will be formed \citep[e.g.,][]{Bundy05, Vergani08}. This informs us that the manner in which baryons flow in and flow out of galaxies, a process called the baryon cycle, depends on the mass of the dark matter halo \citep[e.g.,][]{Katz03, keres05, keres09, birnboim03, Dekel06, Dekel09apj}. The baryon cycle is thought to be the governing process that manifests in the stellar mass-metallicity relationship \citep[e.g.][]{Tremonti04, Mannucci10, Bothwell13}, the stellar mass to halo mass function \citep[e.g.,][]{Kravtsov14}, and the halo mass-star formation rate relationship \citep[e.g.,][]{behroozi13}. Therefore, understanding the baryon cycle and how it affects galaxy formation and evolution is an important part of understanding how galaxies evolve.

Accreting gas falling into galaxies from the intergalactic medium (IGM) passes through the circumgalactic medium (CGM), the metal-enriched gaseous structures surrounding galaxies that act as an interface between the IGM and the interstellar medium (ISM). The general picture is that there are two modes of IGM accretion through the CGM, cold-mode and hot-mode \citep[e.g.,][]{white78, white91, keres09, vandevoort+schaye12}. In lower mass halos, the cooling time is shorter than the dynamical time; this cold-mode accreting gas is able to accrete into the ISM and form stars \citep[e.g.,][]{birnboim03, Dekel06}. In higher mass halos, the cooling time is longer than the dynamical time and the inflowing gas is shock heated. In this hot-mode accretion, much of the gas remains too hot to accrete into the ISM and form stars \citep[e.g.,][]{Fall80, Mo98}, though recent findings indicate that some can accrete into the galaxy \citep{hafen20}. Whereas cold-mode accreting gas passes through the CGM, hot-mode accreting gas remains in the CGM \citep[e.g.,][]{ford13}.  Additionally, large scale galactic outflows from supernovae, stellar winds, and radiation originating in the ISM can propel gas out of the galaxy \citep[e.g.,][]{oppenheimer08}. 
Through a combination of inflows and outflows of gas cycling through the CGM, the star formation rate and stellar content of galaxies is regulated \citep{dave11a, lilly-bathtub}. 

While the details of the CGM's role on the galaxy properties are not yet well known, 
understanding the gas composition, distribution, and dynamics of the CGM is an important aspect of understanding how galaxies evolve \citep[see][for a review]{TumlinsonReview17, kacprzak17}.  

Hydrodynamic cosmological simulations play an important role in understanding how the baryon cycle governs the evolution of galaxies. In the process of modeling the baryon cycle, simulations have been generally successful at producing observed galaxy properties, such as the distribution of halo masses, stellar mass to halo mass function, mass metallicity relation, Tully-Fisher relation, rotation curves, halo mass to star formation rate, etc.\ \citep[e.g.,][]{Behroozi10, behroozi13, moster13, munshi13, Trujillo15, ceverino14, Agertz15}. 

One of the current challenges is for the simulations to successfully reproduce the observable properties of the baryon cycle.  Because the CGM is observable using the technique of quasar absorption lines, we have a rich database of observations of the CGM gas properties at $z \leq1$ \citep[e.g.,][]{chen01a, chen10a, Kacprzak08, ggk-sims, kcn12, kacprzak15, bordoloi-cosdwarfs, tumlinson11, magiicat2, magiicat1, nielsen17, stocke13, Pointon_2019}.  These  observations can be exploited to inform us how well the simulations reproduce one of the largest, massive, and dynamically active reservoirs of galactic gas. If mock observations of the simulations are able to reproduce the observational data, we gain confidence that the physics in the simulations is capturing the physics of the baryon cycle. On the other hand, if the simulations do not reproduce the observations, we gain insight into the physical processes that require refinement.  The challenge is that any refinements to the physics must not inadvertently undermine the success of the simulation to reproduce the global properties of the galaxies themselves.  

CGM gas absorption traced by the {\OVIdblt} doublet has been studied extensively such that the distribution and kinematics of the absorbing gas are well established in relation to the central galaxies \citep{tumlinson11, stocke13, kacprzak15, nielsen17, Kacprzak_2019}. We have the opportunity to place meaningful constraints on how well simulations successfully reproduce the observed distribution, covering fraction, and kinematics of {\OVI}-bearing CGM gas. 

For example, using mock observations of {\OVI} absorption, \citet{Kacprzak_2019} examined the kinematics of {\OVI} using the VELA simulations of \citet{ceverino14, zolotov15, roca19} and concluded that any kinematic signatures of outflows or inflows are washed out by the overall velocity distribution of the {\OVI} throughout the halo.  Their insights provided a theoretical understanding for the findings  of \citet{nielsen17}, who showed that {\OVI} kinematics were indistinguishable for face-on or edge-on galaxies and for various spatial locations relative to the galaxy projected major axis.

In this paper, we extend these studies by further examining whether the {\it spatial\/} distributions of {\OVI}-absorbing CGM gas in the VELA simulations of \citet{Ceverino10, Ceverino16} are consistent with the observed distributions. In particular, we examine the covering fraction of {\OVI} as a function of both impact parameter (normalized by virial radius) and azimuthal angle (the primary angle between the sky projected major axis of the galaxy and the quasar line of sight). The observational data we compare to were presented by \citet{kacprzak15}, who showed a bimodality in the covering fraction that peaks within $\sim\!30\degree$ of the minor axis and in the range $\sim\!10\degree$--$20\degree$ of the major axis. This bimodality, which is also seen in {\MgII} absorption \citep{bouche12, kacprzak12}, has been interpreted as bi-polar (minor axis) outflowing winds and planar (major axis) accretion  \citep[e.g.,][]{stewart11, bouche12, zabl20}. More generally, using the simulations, we aim to characterize the spatial distribution of {\OVI} absorbing CGM gas with respect to the central galaxy disk and to relate this to the inflowing and outflowing gas kinematics.

In Section \ref{Methodology} we describe the observed sample, the simulations, and experimental design. In Section \ref{Results} we present our comparison of the {\OVI} covering fraction as a function of impact parameter and azimuthal angle for the observed and simulated samples. We investigate interpretations of our findings in Section \ref{Discussion}. Finally, in Section \ref{Conclusions}, we discuss what can be inferred from our findings, and offer concluding remarks. Throughout we adopt an $H_{0} = 70$~{\kms}~Mpc$^{-1}$, $\Omega_{\tM} = 0.3$, $\Omega_{\Lambda} = 0.7$ cosmology.

\section{Methodology}
\label{Methodology}

We use hydrodynamic cosmological simulations to study the spatial distribution of {\OVIdblt} absorption in the CGM of simulated Milky Way-type galaxies at $z\approx1$ using mock quasar sightlines. We focus on the spatial distribution of the covering fraction of {\OVI} absorption and compare the simulation results to an observational sample of {\OVI} absorbing galaxies.

\subsection{{Observed Sample}}
\label{sec:obs-sample}

The observational sample we adopt comprises the 53 galaxies of the Multiphase Galaxy Halos Survey built from our (\textit{Hubble Space Telescope\/} ({\it HST\/}) program (PID~13398), the {\it HST\/} archive, and the literature. Full details of the sample have been presented in \citet[][hereafter, \citetalias{kacprzak15}]{kacprzak15}, \citet{Kacprzak_2019}, \citet{Pointon_2019}, and \citet{NG2019}; we briefly summarize the sample here.

All galaxy-absorber pairs have spectroscopic redshifts, which range from $0.08 \leq  z \leq 0.67$ with a mean redshift of $\langle z \rangle = 0.29$.  These galaxies all lie within an impact parameter of $D=200$~kpc, (projected separation of a background quasar). The galaxies are ``isolated'' in that there are no other similarly-bright galaxies within 100~kpc or within a line-of-sight velocity of $\pm 500$~{\kms}.  All galaxies are imaged by \textit{HST} with either the ACS, WFC3, or WFPC2 instruments.   Galaxy morphological parameters, inclinations, and sky orientations were obtained using the modeling software GIM2D \citep{simard02}. 

The {\OVI} absorption line properties were measured in \textit{HST}/COS G130M and G160M spectra of the background quasars.  The absorption line profiles are presented in \citet{nielsen17}, \citet{Pointon_2019}, and \citet{NG2019}. Of the 53 galaxies, 29 have detected absorption with rest-frame equivalent widths $W_{r}(1031) \geq 0.1$~{\AA} and 24 have upper limits or measured absorption with $W_{r}(1031) <0.1$~{\AA}. Hereafter, following the terminology of \citetalias{kacprzak15} for the computation of the covering fraction, we refer to the former as ``absorbers'' and the latter as ``non-absorbers''.  We will adopt these definitions for the mock absorption measurements from the simulations.

\subsection{Selection of Simulated Galaxies}
\label{Sims}

    \begin{figure*}[t]
   \centering
   \includegraphics[width=0.98\hsize]{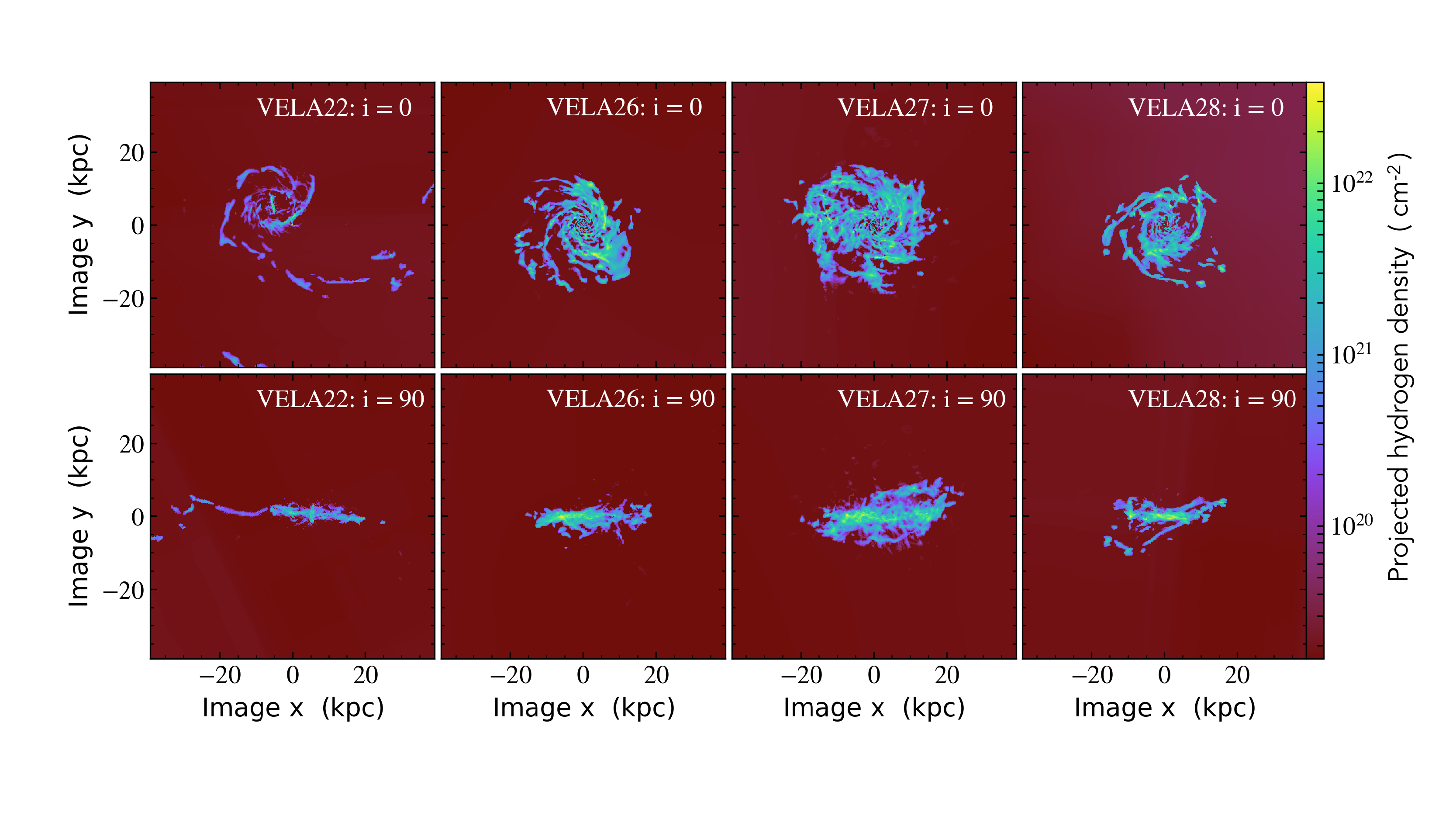}
  \vglue -0.35in \caption{Hydrogen density projection plots of the four VELA galaxies for face-on orientation (top panels) and edge-on orientation (bottom panels) for cool gas ($T \leq 10^{4}$K). }
         \label{PrettyPicture}
       \vglue 0.1in
   \end{figure*}

We adopt a sub-sample of massive galaxies from the VELA simulations \citep{ceverino14, zolotov15}, as shown in Table \ref{table1}. They use the ART code \citep{Kravstov97, Kravstov99_thesis, Kravtsov_2003, ceverino09}. ART combines dark matter $\Lambda$CDM cosmological simulations using an N-body adaptive refinement tree (ART) code and Eularian methods to treat hydrodynamics while employing the zoom-in technique of \citet{Klypin2001}. These simulations can achieve a high-resolution region extending 1–2 Mpc in diameter around simulated galaxies and a cell spatial resolution of approximately 20 pc at z = 1.

The VELA simulations are 20 Mpc on a side. We use post-production smaller boxes that are centered on target galaxies and are roughly four virial radii ($4 R_{vir}$) in diameter. The simulations have a maximum cell resolution of $17$~pc, a minimum stellar particle mass of $10^{3}$~M$_{\odot}$, and a dark matter particle mass of $8 \times  10^{4}$~M$_{\odot}$. This high resolution allows us to resolve the regime in which stellar feedback overcomes radiative cooling \citep{ceverino09}, which results in natural galactic outflows \citep{Ceverino10, Ceverino16}. This allows for a combination of cold flow accretion, mergers, and galactic outflows that results in galaxy formation and evolution proceeding on physically-based principles.

We selected simulated galaxies with halo masses on the order of $11.3 \leq \log M_{vir}/M_{\odot} \leq 11.8$, which is in the range of the sample of observed comparison galaxies. While this may introduce a ``progenitor bias'' due to mass growth of galaxies and the difference in redshift between the observed and simulated sample, our goal was to study galaxies with similar $M_{vir}/M_{\odot}$. We further enforced the condition that the galaxies were isolated from other large galaxies in the simulation volume and that none had experienced a major merger (defined by a mass ratio of 0.3 or greater) for at least 1~Gyr prior to $z\simeq$1. We apply these criteria because the observed comparison galaxies are isolated (see \S~\ref{sec:obs-sample}) and their ``normal'' morphologies suggest none have recently suffered a major merger. 

Ideally, we would prefer to study simulated galaxies between $0.1\leq z\leq 0.7$ to match the redshift range of the observed comparison galaxy sample (with mean $z \simeq 0.3$), but massive VELA galaxies were not evolved below $z\simeq 1$. \citet{Dekel09apj} find that systematic changes in the morphology and mass flux of infalling filaments occurs at $z< 1$ such that the subsequent lack of ``clumps'' carried in these filaments allows enhanced disk stabilization and relaxation at $z<1$. This introduces some concern (which we discuss further in Section~\ref{cov-fac-discuss}), but, as described above, we have attempted to control for this effect by selecting simulated galaxies for which a major merger has not occurred for at least 1~Gyr prior to $z\simeq1$. With regards to global CGM properties, there is no observed evolution of the projected radial profiles of absorbing gas strength \citep{chen12}, which is a favorable observational result with regard to our being limited to $z\simeq1$ simulated galaxies.

Finally, since we aim to study the covering fraction as a function of azimuthal angle with respect to the galaxy disk, we selected simulated galaxies for which we could verify a disk morphology. We define the rotation axis of the galaxy disk to be the angular momentum vector of the cold gas, defined to have $T \leq 10^4$~K, computed in a volume of  $0.1R_{vir}$ around the galaxy center. This angular momentum vector provided the orientation of the galaxy in the simulation box. We then examined sky-projected images of each galaxy's hydrogen number density as a function of galaxy inclination to visually confirm that we could study the galaxies as a function of their ``observed'' inclination. 

Four of the VELA simulated galaxies fit all our selection criteria. Out of these four galaxies, the last major merger was at $z=2.25$, corresponding to $\sim 3$~Gyr prior to $z=1$.  We list the four simulated galaxies and their general properties in Table~\ref{table1}. Sky-projected images of the galaxies are shown in Figure~\ref{PrettyPicture} for both their face on view, $i=0\degree$ (line of sight parallel to the angular momentum vector), and edge-on view $i=90\degree$ (angular momentum vector parallel to the plane of the sky). 

\begin{table}[ht]
\centering
\caption{VELA Galaxy Sample}
\begin{tabular}{c c c c c c} 
\hline\hline 
No. & log(M$_{vir}$/M$_{\odot}$) & log(M$_{*}$/M$_{\odot}$) & R$_{vir}$[kpc] & SFR[M$_{\odot}$~yr$^{-1}$]\\ [0.5ex] 
\hline 
22 &  11.8 & 10.7 & 133 &1.5\\
26 &  11.6 & 10.4 & 112 &1.0 \\
27 &  11.6 & 10.3 & 110 &0.7 \\
28 &  11.3 & 9.9 & 92 &0.2  \\
\hline 
\label{table1}
\end{tabular}
\end{table}

\subsection{Mock Absorption Line Analysis}
\label{Mockspec}

In order to generate the mock quasar spectra and analyze the absorption features, we use the {\sc Mockspec} pipeline developed by C. Churchill \citep{churchill15} and R. Vander Vliet \citep{rachel_thesis}.  {\sc Mockspec} is publicly available on Github\footnote{https://github.com/jrvliet/{\sc Mockspec}}.  A unique capability of {\sc Mockspec} is the ability to identify (and isolate for further study) the gas cells that give rise to significantly detectable absorption. We can therefore isolate, measure, and analyze the properties of the \textit{absorbing} gas, such as density, temperature, metallicity, 3D velocities, dynamic relationship to the simulated galaxy, and 3D spatial location.

Each gas cell in ART has a physical size, $L_{\rm cell}$, and a unique 3D spatial coordinate.  Also recorded are the cell 3D velocity components, hydrogen number density, $n_{\tH}$, kinetic temperature, $T$, and metal mass fraction, $x_{\tM}$.  To obtain the number densities of all ion stages, we perform post-processing equilibrium ionization modeling to obtain the ionization fractions. We use the photo+collisional ionization code {\sc HartRate} \citep[detailed in][]{cwc14}. We adopt solar abundance mass fractions for the individual metals up to zinc \citep{Draine, Asplund09}. This code works best for optically thin, low density gas, making it well suited for studying the low-density CGM ($\log n_{\tH}/{\rm cm}^{-3}  <-1$).   

{\sc HartRate} has been used in previous studies  \citep{cwc1317b, kcn12, churchill15, Kacprzak_2019}. A comparison to the industry-standard ionization code Cloudy \citep{Ferland98,Ferland13} shows that the ionization fractions are in agreement within $\pm 0.05$ across astrophysically applicable densities and temperatures for optically thin gas \citep{cwc14}. All gas cells in post-production boxes (roughly $4R_{vir}$ in diameter centered on the galaxy) are illuminated with the ultraviolet background (UVB) spectrum of \cite{HaardtMadau2005} and the equilibrium solution is obtained. For each cell, {\sc HartRate} records the electron density, ionization and recombination rate coefficients, ionization fractions, and number densities of all ionic stages from hydrogen through zinc.

{\sc Mockspec} generates a user-specified number of "quasar" lines of sight (LOS) distributed through each simulated galaxy. For a given galaxy, the LOS path is defined by its position angle on the plane of the sky, its impact parameter, and the sky-projected inclination of the galaxy. The position angle resides in the range $0^{\circ} \leq \phi \leq 360^{\circ}$, and the plane of the sky is defined as the plane through the galaxy center of mass that is perpendicular to the LOS. The impact parameter is defined as the minimum projected distance between the LOS and the galaxy center of mass, which is equivalent to the distance between the point where the LOS intersects the plane of the sky and the center of mass of the galaxy.  The inclination of the galaxy is defined by the angle between the angular momentum vector and the plane of the sky. 

Observationally, when we speak of the orientation of a given LOS through a galaxy halo, we refer to the galaxy inclination and the LOS azimuthal angle, which is the primary angle ($0\degree\leq \Phi \leq 90\degree$) between the sky-projected major axis of the galaxy and the LOS. We adopt this definition for the mock LOS through the simulated galaxies, where $\Phi = 0\degree$ indicates the major axis and $\Phi=90\degree$ indicates the minor axis. The conversion of the position angle $\phi$ to the azimuthal angle $\Phi$ is a simple rotation and conversion to the primary angle.  The methodology to produce synthetic spectra from quasar sightlines has been outlined in \citet{churchill15} and \citet{rachel_thesis}. Absorption features are detected objectively using the methods described in \cite{churchill00}, which is derived from methods described in \cite{schneider93}.

\subsection{Experimental Design}
\label{ExperimentalDesign}

We aim to compare the simulated sample of {\OVI} galaxy--absorber pairs to the observed sample of {\OVI} galaxy--absorber pairs studied by \citetalias{kacprzak15}, focusing on the covering fraction as a function of impact parameter and azimuthal angle. For the impact parameter comparison in this study, we normalized the impact parameter by the virial radius ($D/R_{vir}$). As we are limited to studying only four simulated galaxies, we must pragmatically adopt the philosophy that {\it many LOS through a few galaxies is equivalent to single LOS through many galaxies}. 

This assumption is observationally supported by the facts that (1) the kinematic distribution of {\OVI}-absorbing gas is uniform around galaxies \citep{nielsen17}, and (2) the spatial bimodality signature is highly significant in a sample of as few as 29 ``absorbing'' galaxies and 24 ``non-absorbing'' galaxies, which implies any individual galaxy is likely to exhibit spatial bimodality of its {\OVI}-absorbing gas. Further, one of our goals is to learn whether individual simulated galaxies that are isolated and have no recent major merger history have {\OVI}-absorbing halos that exhibit covering fraction distributions consistent with the observations. The results will provide insights into the feedback prescriptions, which are physics based. 

In order to control our experiment to match observations as closely as is plausible, we create ``mock samples'' that emulate the number of galaxy--absorber pairs and the distributions of $D/R_{vir}$ and orientation of the LOS in the observed sample.  As described below, we account for variations in the realizations of each mock sample and estimate uncertainties in the covering fractions, by averaging thousands of mock samples.

We began by creating a grid of LOS through the four simulated galaxies. For each galaxy, we ran {\sc Mockspec} at ten inclination angles ranging from $i=0\degree$ (face-on) to $i=90\degree$ (edge-on) in steps of $10\degree$. For each galaxy at each inclination angle, we configured {\sc Mockspec} to generate 1000 LOS with a random distribution of impact parameters between $0 \leq D/R_{vir} \leq 1.5$ and a random distribution of position angles on the sky between $0\degree \leq \phi \leq 360\degree$. We convert the position angles to the primary azimuthal angles, which range between $\Phi = 0\degree$ (probing along the projected major axis) and $\Phi=90\degree$ (probing along the projected minor axis).  This resulted in a library of 40,000 LOS, with each LOS defined by which galaxy it probes, the galaxy inclination, the LOS impact parameter, and LOS azimuthal angle.  

For this study, we chose to normalize the true impact parameter by the virial radius for two reasons. First, we are able to obtain a view of the observed and mock samples normalized to the galaxy mass and mean dark-matter halo overdensity, which is a more commonly adopted view for the theoretical perspective.  Second, the observed sample has a mean of $\langle z \rangle \sim 0.3$, whereas the simulated galaxies are at $z \simeq 1$, having not been evolved to lower redshift. As the galaxies span a range of redshifts and masses, normalizing by virial radius places both the observations and simulations on a comparable scale. Given that the virial radii of galaxies grow substantially from $z = 1$ to $z\sim 0$ \citep[e.g.,][]{bryan98}, it is beneficial to do an analysis that normalizes for galaxy mass and mean halo overdensity, as there is evidence, at least for {\MgII} absorption, that the extent of the CGM is self-similar with halo mass \citep[e.g.,][]{churchill13}.

It is not entirely possible to normalize out the evolution in the CGM gas ionization conditions that would track the evolution in the UVB from $z=1$ to $z=0$, as the mean intensity of hydrogen ionizing photons decreased by roughly an order of magnitude over this range \citep{HaardtMadau2005}.  This can change the relative fraction of photoionized gas to collisionally ionized gas on the {\OVI} phase.  We note that the observed sample spans the approximate redshift range $0.1 \leq z \leq 0.7$, with a mean of $z \simeq 0.3$.   For $z \simeq 0.3$, \citet{Oppenheimer16} showed that the distribution of photoionized to collisionally ionized gas is halo mass dependent. In our halo mass range for the simulations ($10^{11.3} \leq \log M_h/M_{\odot} \leq 10^{11.8}$), they find that the {\OVI} arises primarily in photoionized gas. 

Is it possible that this relative balance of photoionized to collisionally ionized {\OVI}-bearing CGM gas would be significantly altered at $z=1$, the epoch of our simulated galaxies? Further insights are provided by \citet{roca19}, who show that the fraction of photoionized to collisionally ionized gas has both a mass and redshift dependence (see their Figure 14); the amplitude in the trends of the ratio are no more than $\Delta f({\rm phot/coll}) \simeq \pm 0.1$ from $z=1$ to $z=0$ with substantial overlap between galaxies in our mass range.  In summary, though it is difficult to definitively assess the effects of mass growth and UVB evolution in the VELA simulations without running them to lower redshift, the findings of \citet{Oppenheimer16} and \citet{roca19} are suggestive that the effect is not substantial.

 \begin{figure}[htb]
   \centering 
   \includegraphics[width=.92\hsize]{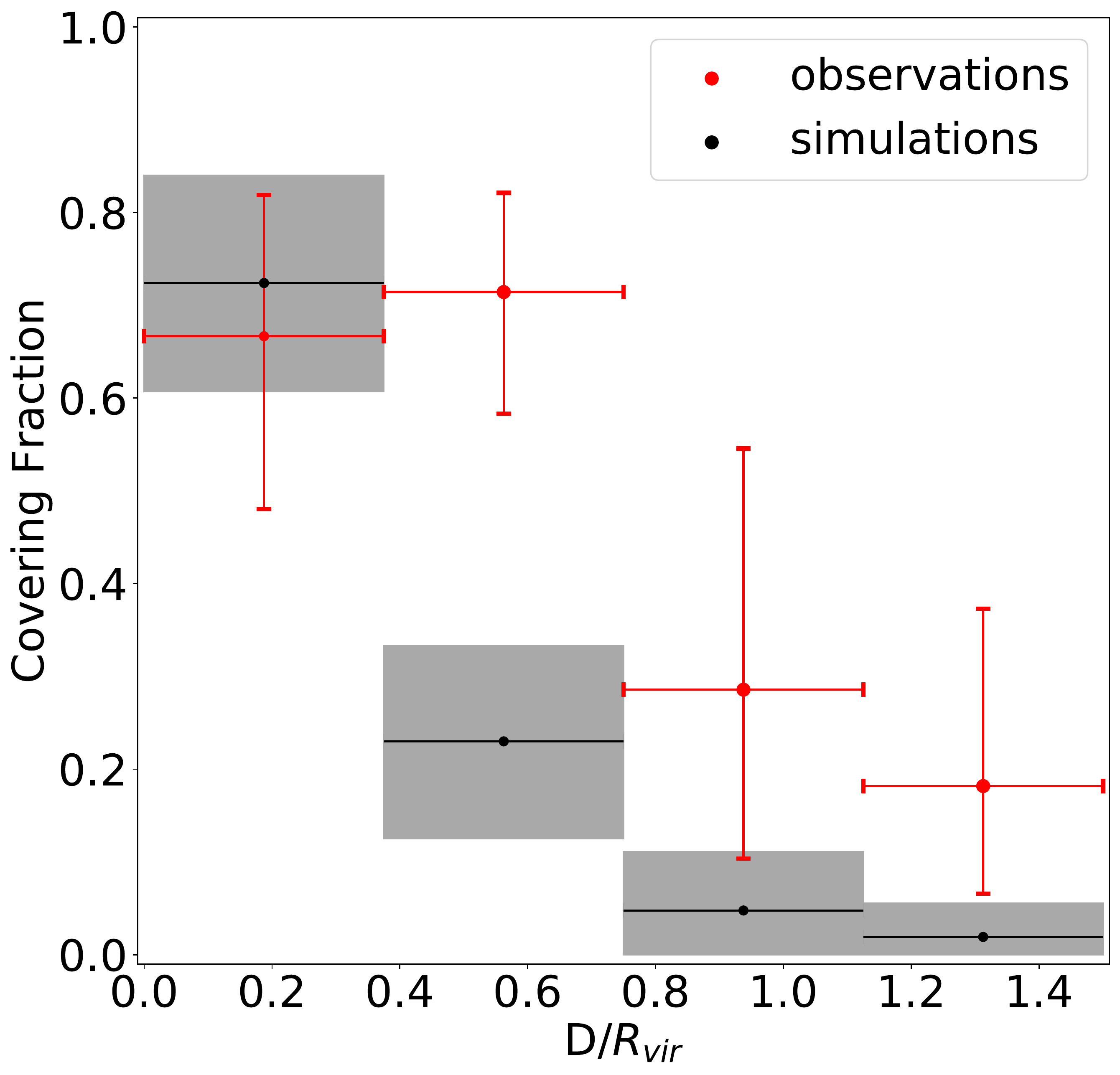}
  \caption{The {\OVI} absorption covering fraction for $W_r(1031) \geq 0.1$~{\AA} (corresponding to log(N$_{\tOVI}/cm^{-2}) \approx 14 $) as a function of impact parameter, $D/R_{vir}$, for $D/R_{vir} \leq 1.5$. Red points are the observed values, similar to those in \citepalias{kacprzak15}, but normalized by virial radius.
  The black points and shaded areas are the means and the $1~\sigma$ confidence levels from the 15,000 mock samples of the simulated galaxies.}
  \label{cf_D}
\end{figure}

For each LOS, {\sc Mockspec} generates a synthetic spectrum covering the {\OVIdblt} absorption.  As the observed {\OVI} absorption properties were measured with the {\it HST}/COS G130M and G160M gratings, we instructed {\sc Mockspec} to generate absorption line spectra with the dispersion and pixel sampling properties of these gratings and the COS FUV CCDs.  We adopted a Poisson fluctuation plus read-noise model \citep[see][]{churchill15} and adopted a signal-to-noise ratio of $S/N=30$ per pixel in the continuum.  This $S/N$ ratio corresponds to a 3~$\sigma$ equivalent width detection limit of $W_r \sim 0.05$~{\AA}. {\sc Mockspec} analyzes the absorption spectra and measures the rest-frame equivalent widths, $W_r(1031)$, and various other quantities that we discard for this study.  Recall that our definition of an {\OVI} ``absorber'' is a detection of $W_r(1031) \geq 0.1$~{\AA}, whereas a ``non-absorber'' is defined by $W_r(1031) < 0.1$~{\AA}, which includes systems not detected to the limit of the spectroscopic data.  All absorption lines generated by {\sc Mockspec} are unblended and no additional noise sources are added.
 
We draw from this library of 40,000 unique LOS to create our mock samples.  Recall that the observed sample comprises 53 galaxies having a distribution of galaxy inclinations and LOS impact parameters and azimuthal angles.  Thus, a given mock sample is created by drawing 53 LOS from the library such that the distribution of inclination angles, impact parameters, and azimuthal angles are consistent with the observed sample.

For each mock sample, we calculated the covering fraction as a function of $D/R_{vir}$ and azimuthal angle. For azimuthal angle, we used the binning employed by \citetalias{kacprzak15}. To account for variations in the covering fraction between realizations of mock samples, we generated and analyzed 15,000 mock samples and compute the mean covering fraction and the $1~\sigma$ confidence intervals of the mean. This interval was determined from the cumulative distribution function (CDF) of all 15,000 realizations.  The upper and lower confidence levels correspond to the 0.8413 and 0.1587 fractional areas under the CDF. 

\section{Covering Fractions}
\label{Results}

 \begin{figure}[t]
   \centering
   \includegraphics[width=0.92\hsize]{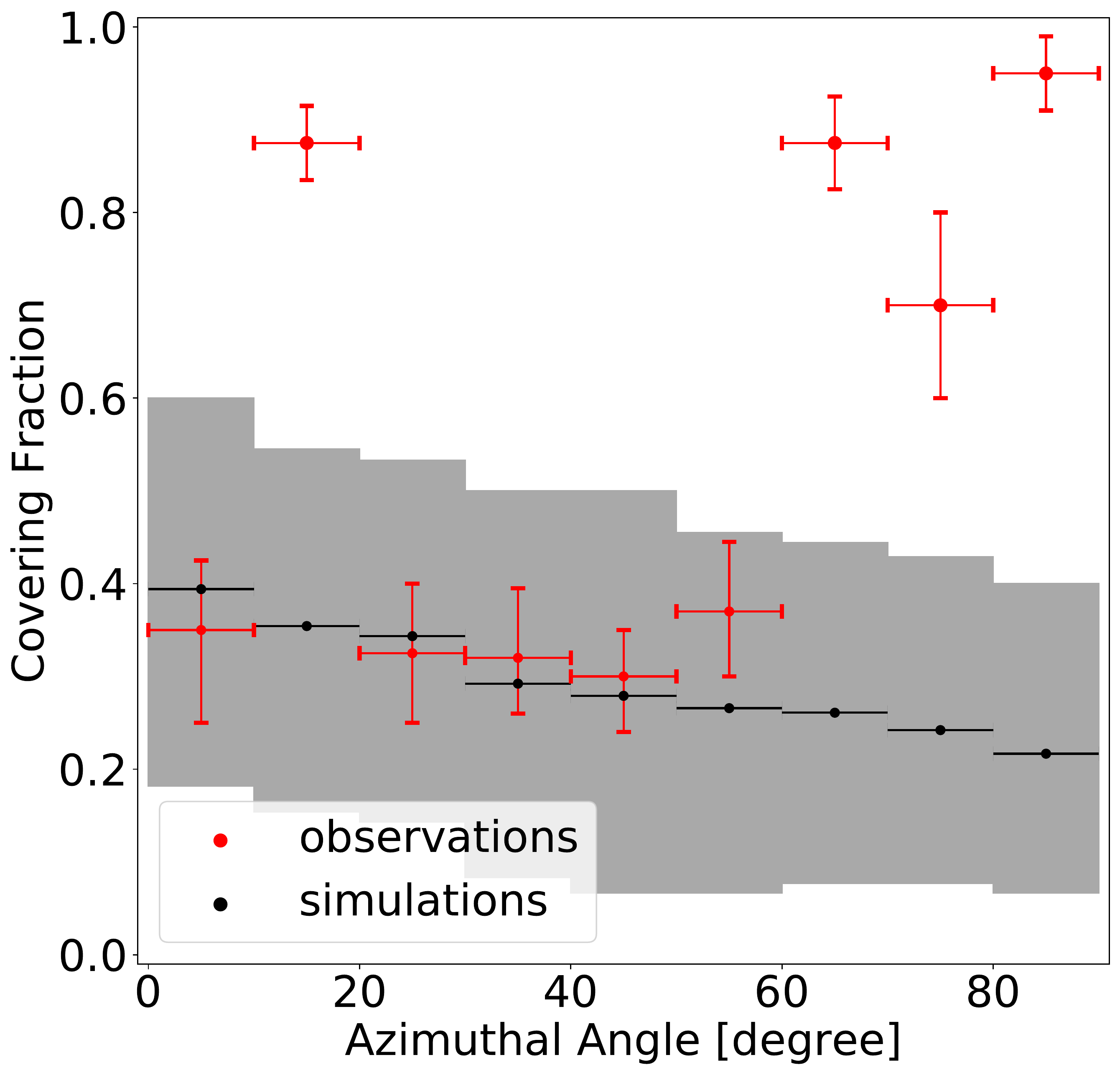}
      \caption{The {\OVI} absorption covering fraction for $W_r(1031) \geq 0.1$~{\AA} as a function of azimuthal angle. Red points are the observed values as presented in $10\degree$ bins \citepalias{kacprzak15}.  The black points and shaded areas are the means and the $1~\sigma$ confidence levels from the 15,000 mock samples of the simulated galaxies. }
         \label{cf_phi}
   \end{figure}

In this section we directly compare our mock samples of {\OVI} galaxy--absorber pairs to the observed sample of \citetalias{kacprzak15}. In Figure~\ref{cf_D}, we present the {\OVI} covering fractions as a function of impact parameter normalized by virial radius for both the observed and mock samples.  We remind the reader that, per the definition of \citetalias{kacprzak15}, an absorber is defined to have $W_r(1031) \geq 0.1$~{\AA} and a non-absorber is defined by $W_r(1031) < 0.1$~{\AA}, which includes systems not detected.  Further, recall that impact parameter in this context is the unitless quantity $D/R_{vir}$, which we employed for the reasons stated in Section \ref{ExperimentalDesign}.

In Figure~\ref{cf_D}, black points are the means of 15,000 mock samples and the grey shaded regions are the $1~\sigma$ confidence intervals of the means.  The mock sample covering fractions are in good agreement with the observed covering fractions in the inner halo ($D/R_{vir} \leq 0.375$).  However, the covering fraction of the mock sample indicates an underproduction of {\OVI} absorbing gas in the projected region $0.375 < D/R_{vir} \leq 0.75$.  Accounting for the uncertainties in the observed covering fractions for $D/R_{vir} > 0.75$, there is greater agreement in the outer halo, where the simulations reproduce the observed covering fraction within $1.1\sigma$\footnote{We account for the multiple realizations from the numerous realizations of the mock sample experiment and the fact that the observed value is to be interpreted as a single realization. Thus, we quote the number of observed $\sigma$ at which the ``uncertainty'' in the simulated mean value resides.}.
Whereas the observations would indicate a clear rapid decline in the {\OVI} covering fraction from $\sim\! 70$\% to $\sim\!20$\% at $D/R_{vir} \sim 0.7$, the VELA simulations yield a rapid decline at $D/R_{vir} \sim 0.4$. 

In Figure~\ref{cf_phi}, we present the comparison of the {\OVI} covering fraction as a function of azimuthal angle for the observed and mock samples.  Similarly to Figure~\ref{cf_D}, black points are the mean of 15,000 mock samples and the grey shaded regions are the $1~\sigma$ confidence intervals of the means.  The $1~\sigma$ confidence intervals suggest a broad distribution of covering fractions are represented in each azimuthal bin. Recall that the mock samples are drawn from the distribution of inclination angles and impact parameters from the observed sample.

The mean covering fraction of the mock samples is remarkably consistent with the observed sample for intermediate azimuthal angles ($20\degree < \Phi \leq 60\degree$). However, the mock samples do not indicate a bimodal distribution in the {\OVI} absorbing CGM of the simulations.  Near the major axis ($10\degree < \Phi \leq 20\degree$) and the minor axis ($\Phi > 60\degree$) the mock samples significantly under-produce {\OVI} compared to the observations. In these spatial regions, the simulated galaxy covering fractions are roughly 25\% of the observed values. Overall, in Figure~\ref{cf_phi}, we see a trend in the mock samples of decreasing covering fraction with increasing azimuthal angle; there is a weak trend for a slightly higher covering fraction of {\OVI} along the major axis of the simulated galaxies.

\section{Analysis and Discussion}
\label{Discussion}

Taken at face value, our analysis reveals that the VELA simulated galaxies exhibit an {\OVI}-absorbing CGM that (1) extends to roughly half of the impact parameter relative to the virial radius, and (2) is more spatially symmetric than that of real galaxies. 
In order to improve our insights into the simulated galaxies, we further investigate the physical properties of the OVI-absorbing gas such as the distribution of densities, temperatures, ionization fractions, metallicity, kinematics, and spatial locations in the simulated galaxies.

\subsection{CGM Gas Phases}

\begin{figure*}[th]
 \begin{minipage}[b]{0.5\textwidth}
  \centering  \includegraphics[width=.92\hsize]{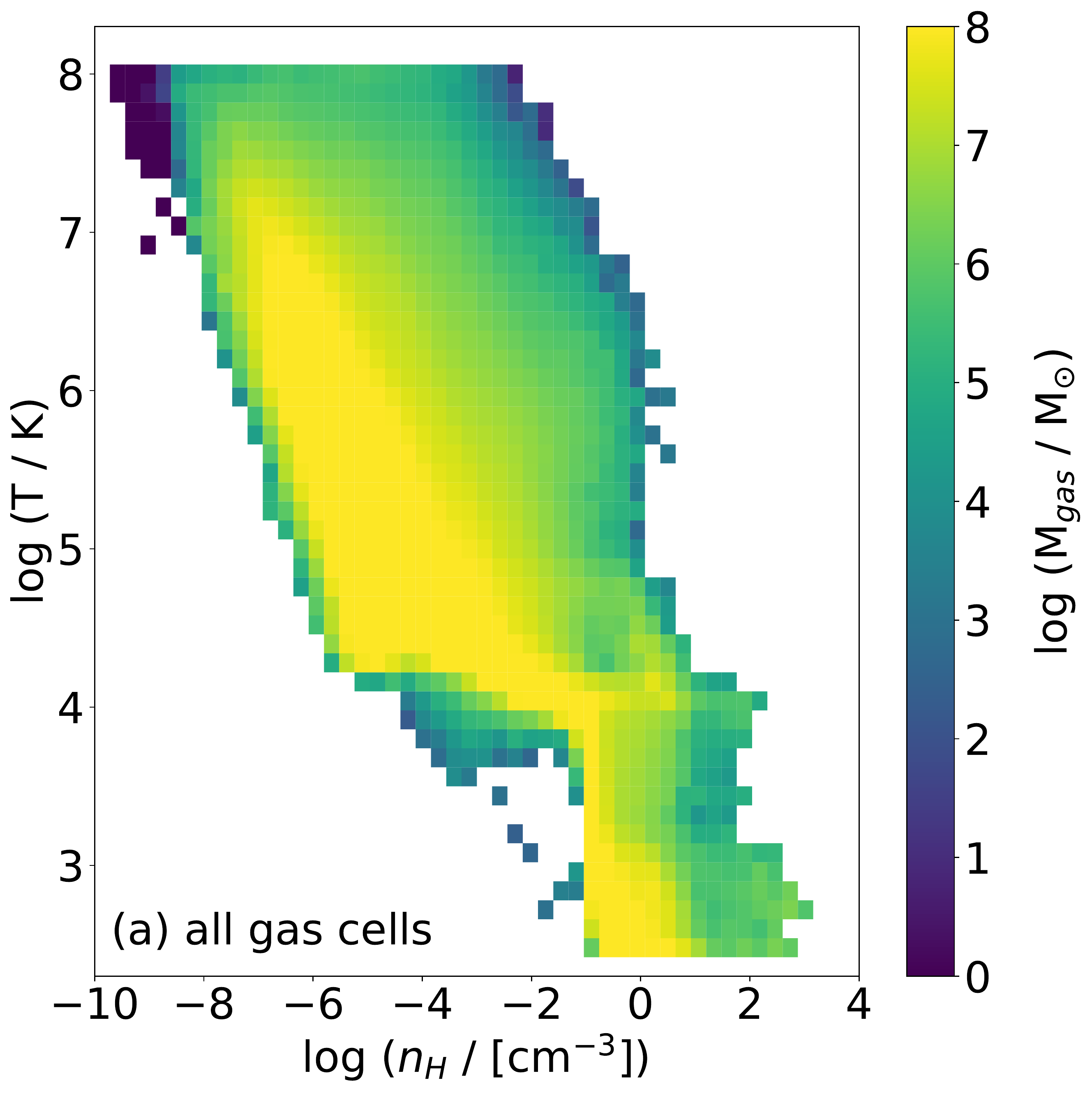}
 \end{minipage}
  \hfill
 \begin{minipage}[b]{0.5125\textwidth}
   \centering  \includegraphics[width=.92\hsize]{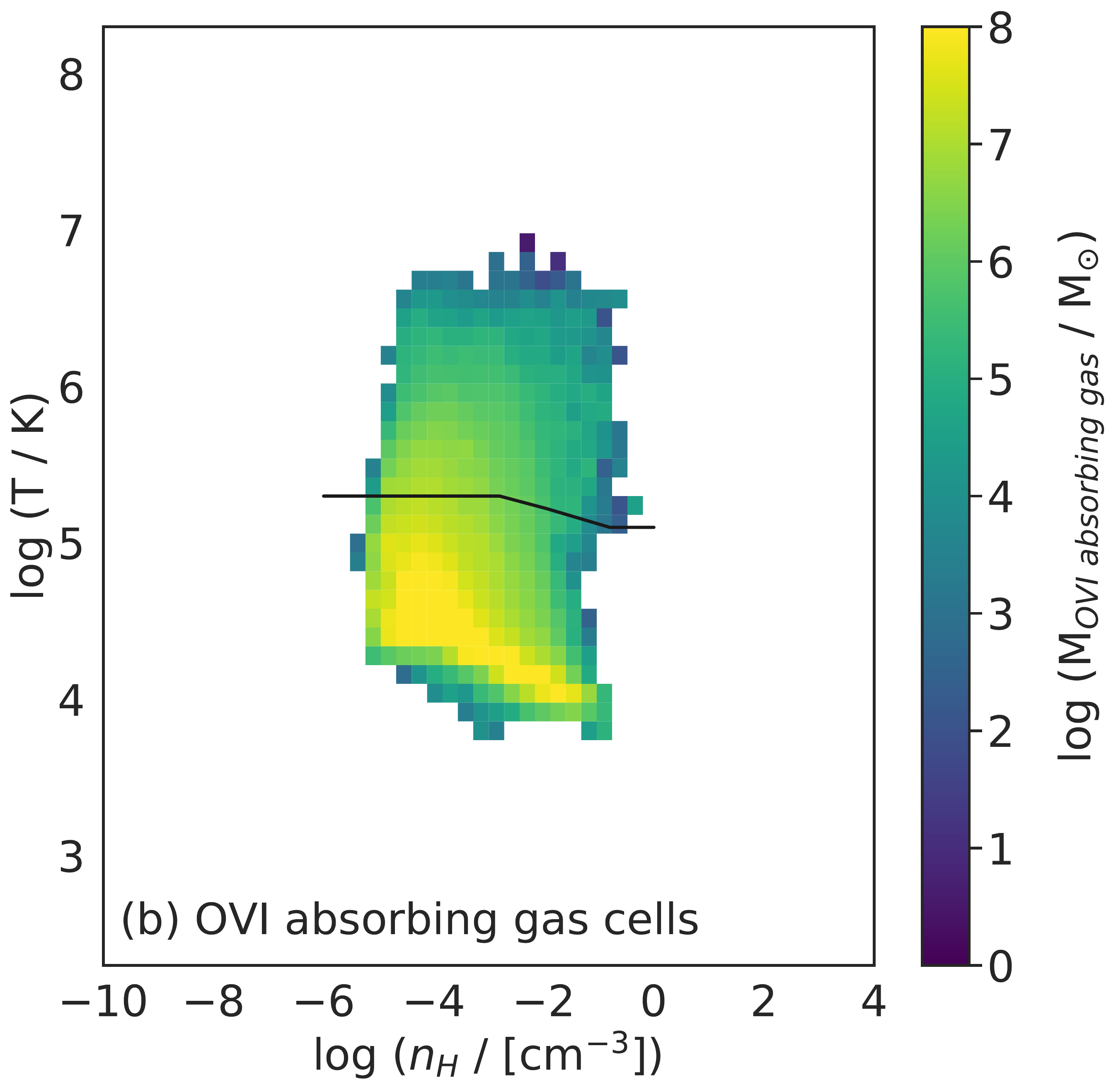}
 \end{minipage}
  \caption{(a) The mass distribution on the $n_{\tH}$--$T$ phase diagram of all gas cells enclosed within the four galaxies. (b) The same as (a), but for the gas cells selected by {\OVI} absorption from all sightlines that probe the four galaxies. Black curve in (b) represents the equilibrium temperature of photoionized gas as a function of $n_{\tH}$ from \citet{Strawn20}; {\OVI} gas below that line is photoionized and {\OVI} gas above that line is collisionally ionized. Note that the pencil beam mock sightlines do not probe the full {\it volume\/} of the CGM, so only the absorbing gas in cells pierced by sightlines can be included.}
  \label{phase}
\end{figure*}

We examined properties of the gas cells that contributed to {\OVI} absorption lines with $W_r(1031) \geq 0.1$~{\AA} in the mock spectra; we hereafter call these "absorbing cells". The details of how absorbing gas cells are selected are described in \citet{churchill15} and \citet{rachel_thesis}.  In brief, these cells contribute more than 5\% to the measured equivalent width of the {\OVI}~$\lambda 1031$ absorption line.  Furthermore, to more broadly characterize the CGM of the VELA galaxies, we compared the absorbing cell properties to the properties of all gas cells within a $\sim 1R_{vir}$ region of the simulated CGM.  

The gas mass of a cell is computed from
\begin{equation}
M_{\rm c} =  \frac{n_{\tH} m_{\tH}} {x_{\tH}} L_{\rm c}^3 \, , 
\label{mass}
\end{equation}
where $n_{\tH}$ is the number density of hydrogen, $m_{\tH}$ is the mass of hydrogen, $x_{\tH}$ is the hydrogen mass fraction, and $L_{\rm c}$ is the length of the cell face.  Employing conservation of mass fractions, $x_{\tH}+x_{\tHe}+x_{\tM}=1$, we estimate $x_{\tH} = {(1-x_{\tM})}/{(1+r)}$, 
where $r = x_{\tHe}/x_{\tH}$ is the ratio of the mass fraction of helium to hydrogen. The value of $r$ ranges from 0.3334 for primordial mass fractions to 0.3366 for solar mass fractions \citep{Lodder19}; this narrow range propagates to a 0.2\% difference in in our estimated cell masses.  We thus adopt a single value of $r=0.335$ for all cells.  This value is appropriate for metal mass fractions of roughly $\sim 0.1$ to 0.01 solar, a value typical of the cells in the simulations.

In Figure~\ref{phase}(a), we show the CGM gas mass distribution on the $n_{\tH}$--$T$ gas phase diagram for all of the gas enclosed within a $1R_{vir}$ volume, where we have co-added all four VELA galaxies. The temperature floor and ceiling of the simulations are $\log T/$K$~ = 2.4$ and $\log T/$K$~ = 8$, respectively. In Figure~\ref{phase}(b), we show the mass distribution of CGM gas selected by {\OVI} absorption drawn from the catalog of 40,000 sightlines.  All four simulated galaxies are co-added, and gas cells in a given galaxy that were probed by multiple LOS (a rare condition) were counted only once. We emphasize that for the absorbing gas, we do not sample the full volume of the CGM, but only the cells along the LOS.

In general terms, for collisionally ionized {\OVI}, we expect gas temperature range $5 \leq \log T/$K$~ \leq 6$ with hydrogen number density in the range $-3 \leq \log n_{\tH}/{\rm cm}^{-3}  \leq -1$, whereas for photoionized {\OVI} absorbing gas, we expect the ranges $4 \leq \log T/$K$~ \leq 4.5$ and $-4.5 \leq \log n_{\tH}/{\rm cm}^{-3}  \leq -3.5$ \citep[e.g.,][]{sutherland93, bergeron05, rachel_thesis}.  The majority of the mass of the gas selected by {\OVI} absorption is warm ($\log T/$K$~\simeq 4.2$), moderately low-density ($\log n_{\tH}/{\rm cm}^{-3} \approx -4$) gas, which is consistent with photoionized {\OVI}. 

As computed from Eq.~\ref{mass}, the average mass of gas in the CGM of the simulated galaxies within $1R_{vir}$,  is $\log M_{CGM}/M_{\odot} = 10.5$, which is comparable to the average stellar mass of $\log M_{\star}/M_{\odot} \simeq 10.4$. Using Figure~\ref{phase}(b) as a guide, if we crudely assume that {\OVI} absorption arises in all CGM gas in the phase defined by $4 \leq \log T/$K$~ \leq 6.5$ and $-5 \leq \log n_{\tH}/{\rm cm}^{-3} \leq -1$, as this is the temperature and number density range of hydrogen which produces 98\% of the {\OVI} mass, the average mass of gas selected by {\OVI} absorption is $\log M_{\hbox{\tiny OVI}}/M_{\odot} = 10.4$.  This should be considered an upper limit, but it indicates that a significant amount of gas mass in the CGM is in the phase that would be detected in {\OVI} absorption. 

Conversely, comparing to the non-absorbing CGM gas, which we define to be all gas with $\log T/$K$~ \geq 4$ that resides outside the phase region defined by $4 \leq \log T/$K$~ \leq 6.5$ and $-5 \leq \log n_{\tH}/{\rm cm}^{-3}  \leq -1$ (which we used to approximate the {\OVI}-absorbing gas), we find that the average volume of {\OVI}-absorbing gas in the CGM of a simulated galaxy is $\log V_{CGM}/$ kpc$^3 = 6.0$.  This is a factor of five smaller than the volume filled by non-absorbing gas ($\log V_{\hbox{\tiny OVI}}/$kpc$^3 = 6.7$). The statistically equivalent radius of a sphere encompassing the absorbing gas is $\simeq 60$~kpc, which corresponds to $\simeq 55$\% of the average virial radius of the simulated galaxies ($\bar{R}_{vir} = 112$~kpc). Alternatively, the equivalent radius for the non-absorbing gas is $\simeq 105$~kpc, which corresponds to $\simeq 95$\% of the average virial radius of the simulated galaxies.  The relatively small volume of {\OVI}-absorbing gas is consistent with an expectation of a relatively low covering fraction in the simulated VELA galaxies. 

As with our mass estimates, the volume of the {\OVI} absorbing gas that we estimate here should be viewed as an upper limit because we make these calculations employing a region of the gas-phase diagram defined by $4 \leq \log T/$K$~ \leq 6.5$ and $-5 \leq \log n_{\tH}/{\rm cm}^{-3}  \leq -1$. This is done since LOS, being pencil beam probes, do not sample the full volume of the CGM. It is likely that not all gas in the defined phase range gives rise to detectable {\OVI} absorption.

A favorable combination of ionization conditions and chemical enrichment increase the likelihood that CGM gas will give rise to {\OVI} absorption.  Again, approximating the gas-phase region $4 \leq \log T/$K$~ \leq 6.5$ and $-5 \leq \log n_{\tH}/{\rm cm}^{-3}  \leq -1$ for {\OVI}-absorbing gas and all other gas on the phase diagram with $\log T/$K$~ > 4$ as non-absorbing CGM gas, we find that the distribution of {\OVI} ionization fraction of absorbing gas is highly peaked with a single mode at $\log f_{\tOVI} = -1$ and a flat tail out to $\log f_{\tOVI} = -5$.  On the other hand, the distribution for non-absorbing gas is normally-distributed with a single mode at $\log f_{\tOVI} = -7$ and a HWHM of $\pm 2.5$ dex. 

The distribution of metallicity for the absorbing gas is highly peaked, comprising a single mode at $\log Z/Z_{\odot}= -0.7$ with a high metallicity tail up to the solar value and an extended low metallicity tail down to $\log Z/Z_{\odot}= -2$.  The non-absorbing gas has a bimodal distribution, with a highly peaked and extremely narrow mode at $\log Z/Z_{\odot}= +0.1$, and a broader normally-distributed peak at $\log Z/Z_{\odot}= -0.7$.  The high metallicity mode is primarily very hot and diffuse gas ($\log T/$K$~ \geq 6.5$ and $\log n_{\tH}/{\rm cm}^{-3}  \leq -6$). This component of the CGM may be highly metal-enriched, but it comprises less than 5\% of the gas mass within $1R_{vir}$ of the galaxy.  In these simulations, it is the very hottest gas that carries the markings of recent processing through stellar feedback.

\subsection{Gas Kinematics: Infall and Outflow}
\label{kinematics}

As inflowing and outflowing gas through the CGM are two major components of the baryon cycle, it is of interest to investigate the relative contributions of these components to the covering fraction. In particular, insights into the rapid drop in the covering fractions at $0.375 < D/R_{vir} \leq 0.75 $ in the mock sample (see Figure~\ref{cf_D}) might be connected to the relative contribution of inflowing versus outflowing {\OVI}-absorbing gas as a function of $D/R_{vir}$. To quantify the sky-projected contributions of these two kinematic components, we examined the {\OVI}-absorbing gas mass surface density of {\OVI} absorption-selected gas, ${\Sigma} ({\OVI}$-absorbing gas). 

For a given sky-projected annular region of impact parameter, $D_{\rm in} \leq D \leq D_{\rm out}$, having area $A$, the {\OVI}-absorbing gas mass surface density is 
 \begin{equation}
{\Sigma} ({\OVI}) = 
\frac{M_{\hbox{\tiny OVI}}}{A} \\ 
=  \frac{m_{\hbox{\tiny O}}}{2\pi\left( D^2_{\rm out} - D^2_{\rm in} \right)} 
\sum_{i} n_{i,\hbox{\tiny OVI}} L_{i,{\rm c}}^3 \, , 
\label{OVImass}
\end{equation}
where $m_{\hbox{\tiny O}}$ is the mass of oxygen.  For each sky-projected annular region, the sum is over absorbing cells that reside within the virial radius, where $n_{i,\hbox{\tiny OVI}}$ is the {\OVI} number density in cell $i$ and $L_{i,{\rm c}}$ is the size of cell $i$.  To obtain ${\Sigma} ({\OVI})$ of inflowing absorbing gas, we sum over all cells with a negative radial velocity component, $v_{r} < 0$, and for outflowing absorbing gas we sum over all cells having $v_{r} \geq 0$.  The radial velocity is measured with respect to the galaxy center. We also compute ${\Sigma} ({\OVI})$ for all absorbing cells so we can obtain the total {\OVI} mass surface density.

\begin{figure}[tbh]
   \vglue 0.03in
   \centering
   \includegraphics[width=0.92\hsize]{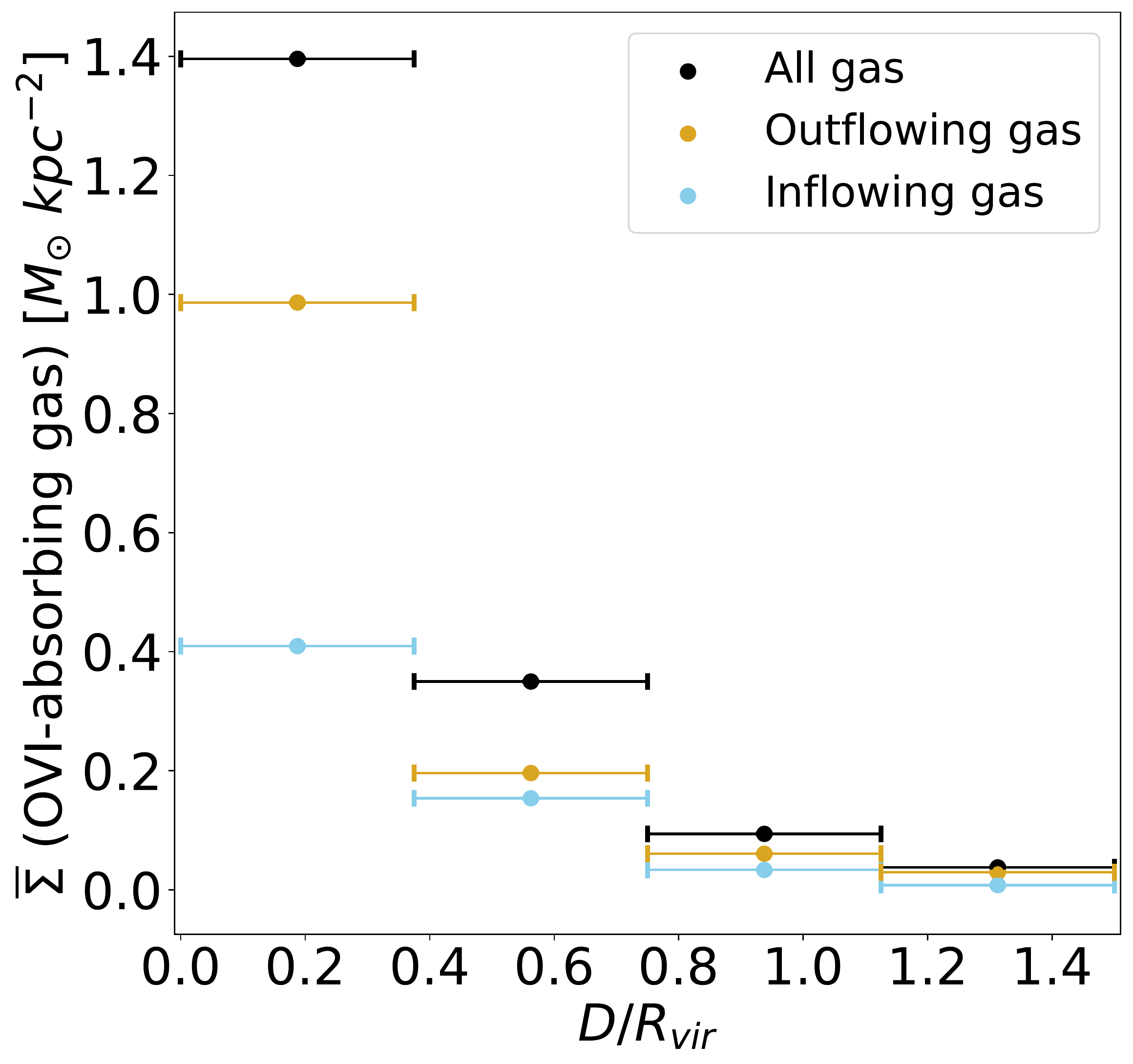}
      \caption{Average {\OVI}-absorbing gas mass surface density of {\OVI} absorption-selected cells, $\bar{\Sigma} ({\OVI}$-absorbing gas), as a function of $D/R_{vir}$ for radially-inflowing {\OVI}-absorbing gas (blue points), radially-outflowing {\OVI}-absorbing gas (gold points), and for all {\OVI} absorbing gas (black points).}
      \vglue -0.1in
         \label{ASD}
   \end{figure}

In Figure~\ref{ASD}, we present the {\it average\/} {\OVI}-absorbing gas mass surface density, $\bar{\Sigma} ({\OVI}$-absorbing gas), as a function of $D/R_{vir}$, using the same binning shown in Figure~\ref{cf_D}.  The average values are obtained by summing over all LOS through all simulation galaxies and dividing by the number of galaxies. 

Overall, the average {\OVI}-absorbing gas mass surface density generally exhibits a decline as function of $D/R_{vir}$ that is qualitatively similar to that of the covering fraction.  In fact, as one increases in $D/R_{vir}$, the relative values of $\bar{\Sigma} ({\OVI}$-absorbing gas) are quantitatively consistent (within uncertainties) with the relative values of the simulated covering fractions from the mock samples presented in Figure~\ref{cf_D}. The total average {\OVI}-absorbing gas mass surface density (black points) is roughly a factor of four higher at $D/R_{vir} \leq 0.375$ as compared to the projected region $0.375 < D/R_{vir} \leq 0.75$, and this matches the ratio of the covering fractions for these regions. That is, in the vicinity of $D/R_{vir} \simeq 0.375$ there is a clear drop off in the sky-projected {\OVI}-absorbing gas mass surface density and this manifests as a similarly proportioned drop off in the covering fraction. This is in contrast with the behavior of the observed covering fraction, which remains virtually constant for $0 \leq D/R_{vir} \leq 0.75$. This suggests that, for the simulations, covering fraction is a good indicator of the average mass surface density of absorbing gas, and vice versa.

For these simulated galaxies, outflowing {\OVI}-absorbing gas is the dominant contributor to the total mass surface density at all $D/R_{vir}$. However, the ratio of outflow to inflow decreases rapidly at $D/R_{vir} \simeq 0.375$.  Beyond this sky-projected region, the ratio of the outflowing and inflowing mass surface densities remains roughly constant. For $D/R_{vir} \leq 0.375$, outflowing gas comprises roughly 70\% of the total {\OVI}-absorbing gas mass surface density, whereas inflowing gas comprises only about 30\%.  Taken at face value, we would infer that roughly 70\% of the {\OVI} covering fraction in the projected region $D/R_{vir} \leq 0.375$ is contributed by outflowing {\OVI}-absorbing gas, whereas the remaining 30\% is contributed by inflowing {\OVI}-absorbing gas.

For $0.375 < D/R_{vir} \leq 0.75$, outflowing and inflowing gas contribute roughly 55\% and 45\%, respectively, to the total {\OVI}-absorbing gas mass surface density.  We may infer that these are the approximate proportions that inflowing and outflowing {\OVI}-absorbing gas contribute to the covering fraction in the mock sample.  Based on the mass surface densities, we find that the contribution to absorption from outflows in the region $0.375 < D/R_{vir} \leq 0.75$ have decreased by a factor of roughly five relative to $D/R_{vir} \leq 0.375$. That is, there is 20\% as much mass surface density of outflowing absorbing gas in the region $0.375 < D/R_{vir} \leq 0.75$ as there is in the region $D/R_{vir} \leq 0.375$. In contrast, the contribution to absorption from inflows in the region $0.375 < D/R_{vir} \leq 0.75$ have decreased by a only factor of two relative to $D/R_{vir} \leq 0.375$. 

In summary, though there is a steady and smooth increase in $\bar{\Sigma} ({\OVI}$-absorbing gas) from the outer to the inner halo for inflowing gas, there is a dramatic increase in $\bar{\Sigma} ({\OVI}$-absorbing gas) in the inner halo for outflowing gas. These trends in the average {\OVI}-absorbing gas mass surface density with $D/R_{vir}$ inform us that outflows dominate the {\OVI} absorption in the inner regions of the simulated CGM.  We infer that the majority of the outflowing gas does not reach the outer regions of the CGM (only $\sim\! 20$\% is reaching to $D/R_{vir} > 0.375$).  We thus attribute the factor of four decrease in the {\OVI} absorption covering fraction at $D/R_{vir} \simeq 0.375$ in the mock sample (see Figure~\ref{cf_D}) to primarily be a consequence of the majority of the outflowing gas being confined to the inner halo in the simulations.

\begin{figure*}[th]
\centering
\includegraphics[width=0.92\textwidth]{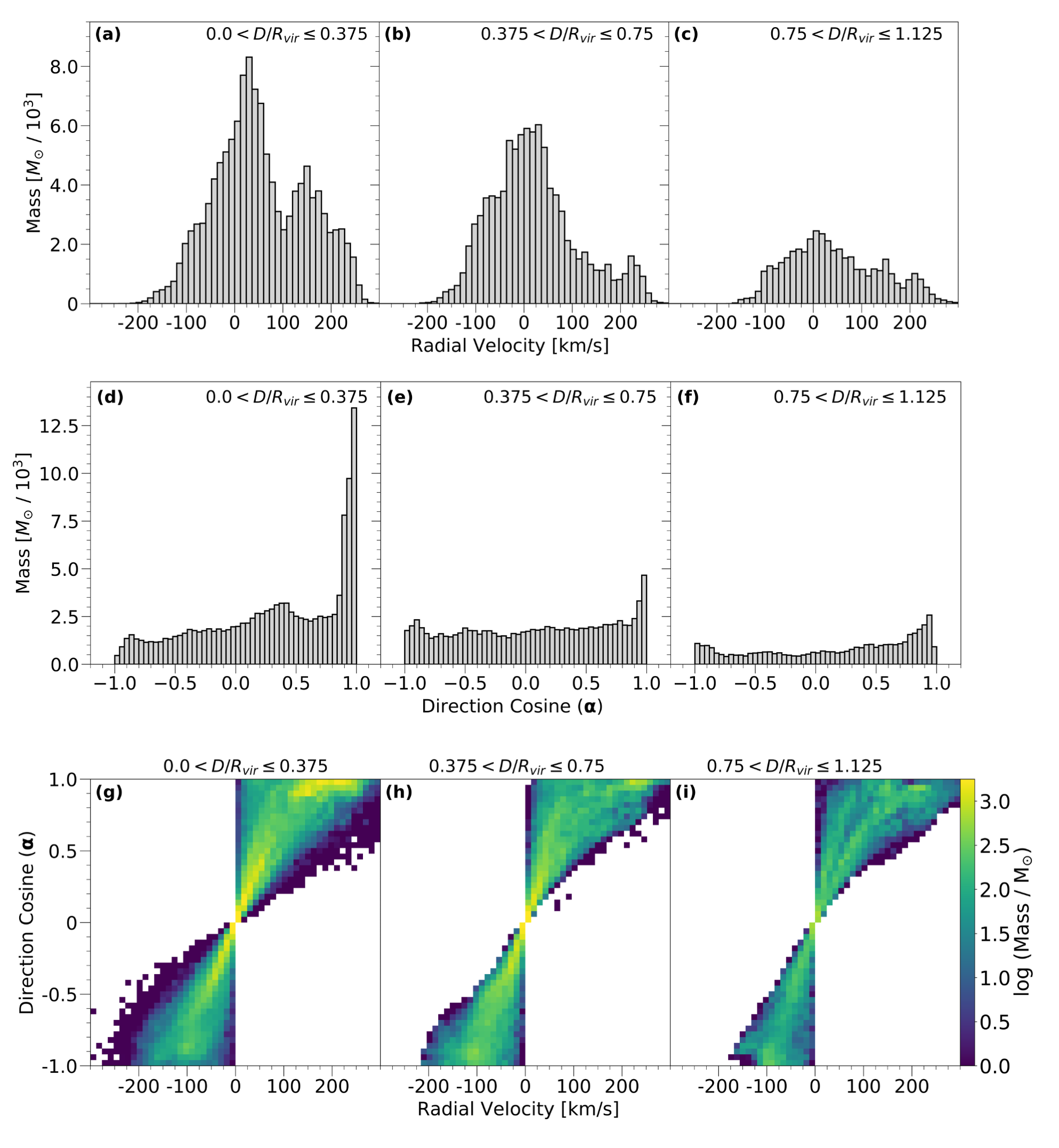}

  \caption{(a,b,c) The {\OVI} mass distribution as a function of radial velocity, $v_r$, for (a) $D/R_{vir}\leq 0.375$, (b) $0.375 < D/R_{vir}\leq 0.75$, and (c) $0.75 < D/R_{vir}\leq 1.125$. (d,e,f) The same as (a,b,c) except for the direction cosine of the radial velocity, $\alpha$. (g,h,i) The 2D {\OVI} mass distribution of radial velocity and the direction cosine of the radial velocity in the same $D/R_{vir}$ bins as presented in panels (a), (b), and (c).}
  \label{comboOVI}
\end{figure*}

To further characterize the inflow and outflow kinematics of the {\OVI}-absorbing gas, we examined the distribution of the {\OVI} mass for {\OVI} absorption-selected cells as a function of both the radial velocity and the direction cosine of the radial component of the velocities.  For a cell with velocity components $v_r$, $v_\theta$, and $v_\phi$, the direction cosine is given by 
\begin{equation}
\alpha = \cos (a) 
= \frac{\mathbf{v}\cdot \hat{\mathbf{v}}_r}{| \mathbf{v} | }
= \frac{v_r}{ \sqrt{v_r^2 + v_\phi^2 + v_\theta^2}} \, , 
\end{equation} 
where $\hat{\mathbf{v}}_r$ is the radial velocity unit vector. The range of $\alpha$ is\, $-1 \leq \alpha \leq +1$, where $\alpha =-1$ indicates pure radial inflow and $\alpha =+1$ indicates pure radial outflow. The {\OVI} mass in each cell is computed from $M_{\rm c}({\OVI}) =  n_{\tO} m_{\hbox{\tiny OVI}} L_{\rm c}^3$. 

In Figures~\ref{comboOVI}(a--c), we present the {\OVI}-absorbing gas mass distribution as a function of radial velocity and, in Figures~\ref{comboOVI}(d--f), we present the  distribution of the radial velocity direction cosine.  The 2D distribution of {\OVI} mass is presented as a function of the radial velocity and the radial velocity direction cosine in Figures~\ref{comboOVI}(g--i). The distributions are plotted for the three sky-projected regions $D/R_{vir}\leq 0.375$, $0.375 < D/R_{vir}\leq 0.75$, and $0.75 < D/R_{vir} \leq 1.125$; which correspond to the inner three regions for which the {\OVI} mass surface density was computed (see Figure~\ref{ASD}).  We did not compute the distributions in the outer-most region ($D/R_{vir} > 1.125$) as both the {\OVI} covering fraction and mass surface density was quite similar to the third region and our focus was to interpret the significant drop seen in $0.375 < D/R_{vir}\leq 0.75$.

For $D/R_{vir}\leq 0.375$, we find two modes in the radial velocity distribution (Figure~\ref{comboOVI}(a)).  The first mode is centered on $v_r \simeq +10$~{\kms} and comprises a roughly equal mixture of inflowing and outflowing {\OVI}-absorbing gas with a dispersion of $\sigma(v_r) \simeq 100$~{\kms}. The second is at $v_r \simeq +180$~{\kms} and comprises outflowing {\OVI}-absorbing gas with a dispersion of $\sigma(v_r) \simeq 50$~{\kms}.  The distribution of the radial velocity direction cosines shows a very strong peak at $\alpha \simeq +1$ (Figure~\ref{comboOVI}(d)), indicating that a substantial proportion of the outflowing absorbing gas is on a pure radial trajectory.  The 2D distribution (Figure~\ref{comboOVI}(g)) clearly reveals a mass peak of high velocity radially outflowing {\OVI}-absorbing gas at $120 \leq v_r \leq 230$~{\kms} and $\alpha \simeq +1$. 

These distributions indicate that the high covering fraction in the region $D/R_{vir}\leq 0.375$ arises due to (1) roughly equal mass of gas both inflowing and outflowing in the velocity range $-100 \leq v_r \leq 100$~{\kms} and (2) a component of {\it radially\/} outflowing gas in the velocity range $120 \leq v_r \leq 230$~{\kms}.  Per our {\OVI} surface mass density analysis, we infer that, by mass, $\sim\! 70$\% of the absorbing gas is associated with the outflow, and this is consistent with the kinematic analysis. Given there is roughly equal mass of inflow and outflow associated with the mass distribution of the low velocity mode ($v_r \sim +10$~{\kms}, see Figure~\ref{comboOVI}(a)), we can now further infer that roughly $\sim\! 40$\% of the total outflow mass is in the form of $v \sim 200$~{\kms} absorbing gas on a highly radial trajectory. 

For $0.375 < D/R_{vir}\leq 0.75$ Figures~\ref{comboOVI}(b,e,h), we find that the total mass of {\OVI}-absorbing gas has decreased and that the radial outflow bimodality is significantly reduced relative to the bulk of absorbing gas in the velocity range $-100 \leq v_r \leq 100$~{\kms}.  

It is apparent that the majority of the high-velocity radial outflow detected in {\OVI} absorption within the projected region $D/R_{vir}\leq 0.375$ has dynamically stalled prior to reaching the region $0.375 < D/R_{vir}\leq 0.75$.  If a majority of the gas had reached this intermediate region and become dynamically mixed with the inflowing and outflowing gas in the velocity range $-100 \leq v_r \leq 100$~{\kms}, we would expect the total gas mass to be higher than what is measured. Consistent with our inference from the {\OVI} mass surface density, we see that roughly 45\% of the absorbing gas is inflowing and 55\% is outflowing.  The kinematics indicate that the 10\% difference is in the form of $v \sim 200$~{\kms} absorbing gas outflowing on a highly radial trajectory.  

In the outer projected region $0.75 < D/R_{vir} \leq 1.125$ (Figure \ref{comboOVI}(c,f,i)), we see a continuation of a drop in total {\OVI} mass for both inflowing and outflowing gas. We also see a remnant of the radial outflow apparent in the extended tail of the radial velocity distribution and the small peak in the direction cosine at $\alpha \sim +1$.  

Corroborating the behavior of the {\OVI} mass surface density with increasing $D/R_{vir}$, the kinematics indicate that the majority of the absorbing gas mass in the inner projected region of the halo of the simulated galaxies is outflowing gas.  Further, the kinematics suggest that the drop in {\OVI} absorbing gas mass with $D/R_{vir}$ is because the majority of the outflowing gas is not making it out past $D/R_{vir}\simeq 0.375$.  There is also significant decrease in the low mass, infalling gas at $D/R_{vir} > 0.75$ here than was present in the lower $D/R_{vir}$ bins.  This suggests a lack of accreting gas in the outer regions of the galaxy, which is potentially a result of the outflowing gas not making it past $D/R_{vir}\simeq 0.375$ and preventing recycled accretion from occurring.

In a study of {\OVI} absorption line kinematics using a slightly larger sample of simulated galaxies from the same VELA suite as our galaxies, \citet{Kacprzak_2019} concluded that the {\OVI}-absorbing outflows do not appear to be present beyond $\sim\!50$ kpc (approximately $D/R_{vir} \sim 0.4$), and that the outflows eventually decelerate and fall back towards the galaxy. Thus, the characteristics we find for our smaller sample of galaxies (based on the selection criteria adopted for this study) are likely the same characteristics of the VELA simulations in general. 

\begin{figure*}[tbh]
 \centering   \includegraphics[width=0.92\textwidth]{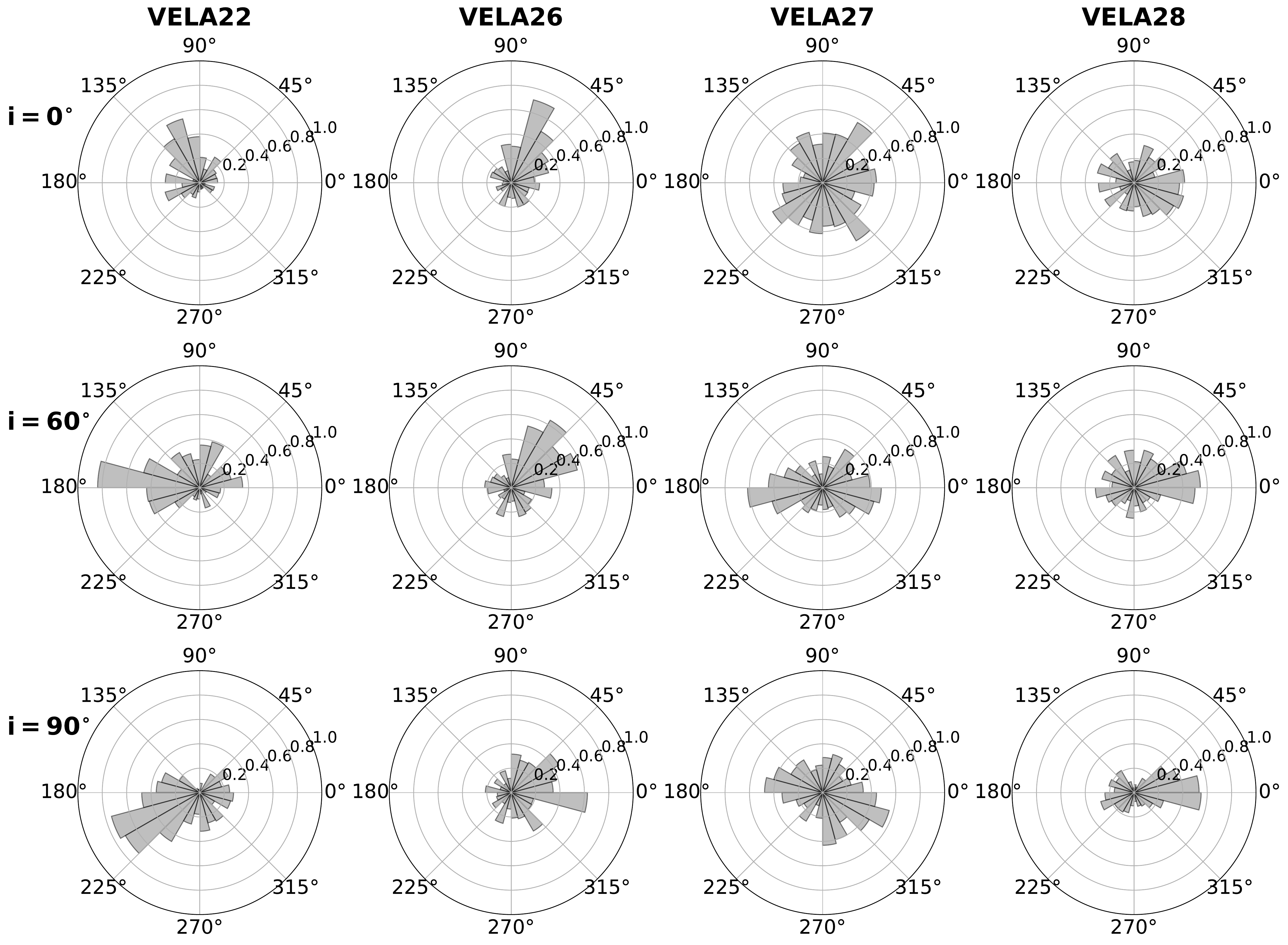}
      \caption{The sky projection of the {\OVI} covering fraction within the projected range $D/R_{vir} \leq 1.5 $ for $W_r(1031) \geq 0.1$~{\AA} as a function of position angle for the simulated galaxies for (upper) $i = 0\degree$ or face-on, (center) $i = 60\degree$, (lower) $i = 90\degree$ or edge-on. Each ring represents covering fraction; taller bars indicate a higher covering fraction at a given position angle.}
         \label{grid_plots}
\end{figure*}

\citet{Strawn20} also study selected simulated galaxies from the same VELA suite we used.  Consistent with our findings, they report {\OVI}-bearing gas is predominantly inflowing at $D/R_{vir} \geq 0.3$.  They further show that this inflowing gas is photoionized, whereas the bulk of the gas halo by volume (not mass) inside  $D/R_{vir} \leq 0.3$ is collisionally ionized.  As we have shown, by mass, the majority of the gas {\it selected by detectable {\OVI} absorption in mock spectra\/} is photoionized (see the phase diagram, Figure \ref{phase}).  For comparison, we further examine the radial distribution of the ionization conditions of our absorption selected gas by examining the phase diagram for $D/R_{vir} < 0.3$ and $D/R_{vir} \geq 0.3$.  We find results consistent with \citet{Strawn20}. The vast majority ($\sim 90$\%) of the inflowing {\OVI}-absorption selected gas mass at $D/R_{vir} \geq 0.3$ is photoionized.  However, for $D/R_{vir} < 0.3$, the fraction of {\OVI}-absorption selected gas mass that is photoionized has decreased to $\sim 60$\%.  Again, the characteristics we find for our smaller sample of galaxies likely reflect the same general characteristics of the VELA simulations.

Considering our findings and those of both \citet{Kacprzak_2019} and \citet{Strawn20}, we suggest that it is the inability of the majority of winds to reach the outer halo that explains the dramatic drop in the {\OVI} covering fraction at $D/R_{vir} > 0.375$ in the mock sample (see Figure~\ref{cf_D}). If the radial outflowing absorbing gas were to reach out to $D/R_{vir} \simeq 0.75$, we might expect a flatter profile of the mass surface density across the region $0 \leq D/R_{vir} \leq 0.75$ and therefore a flatter {\OVI} absorption covering fraction out to $D/R_{vir} \simeq 0.75$ (as measured for the observed sample).

If the simulations provide a window on the processes occurring in real galaxies, then we would be in a position to suggest that the galaxies comprising the observed sample have outflows that reach somewhat beyond $D/R_{vir} = 0.5$, and perhaps as far out as $D/R_{vir} = 0.75$.  In a further analysis of the observed galaxy sample of \citetalias{kacprzak15} and the results of \citet{nielsen17}, \citet{NG2019} yielded evidence for the presence of outflows.
Interestingly, these outflows might be required by the observations to have a clear spatial structure relative to the projected major axis as measured in the modes of the azimuthal distribution of the covering fraction (see Figure~\ref{cf_phi}). 

However, \citet{nielsen17} showed that {\OVI} absorption velocity spreads are similar regardless of galaxy inclination or azimuthal angle and suggest that {\OVI} kinematics is apparently not causally connected to the stellar activity of the central galaxy. \citet{nielsen17} speculate that the observed bimodal distribution in the {\OVI} covering fraction may be due to the gas ionization conditions at intermediate azimuthal angles ($20\degree \leq \Phi \leq 60\degree$) not favoring the {\OVI} phase.  If this is true, the VELA simulations have not captured this characteristic of the CGM. Additionally, for a similar sample of observed galaxies, \citet{NG2019} looks at the {\OVI} gas kinematics relative to the galaxy when accounting for the mass of the host galaxy. Their results suggests that there is a strong halo mass dependence on {\OVI} absorber kinematics and to coax out the signatures of outflows one must account for the halo mass dependence, which we indirectly do in this work by normalizing both the observed and simulated sample by virial radius.

\subsection{Covering Fraction and Galaxy Orientation}
\label{cov-fac-discuss}

As shown in Figure~\ref{cf_phi} and described in \S~\ref{Results}, the covering fraction of the mock sample was in good agreement with that of the observed sample in the azimuthal range $20\degree \leq \Phi \leq 60\degree$, but the mock sample did not yield the azimuthal bimodality found for the observed sample.  The fact that a significant azimuthal bimodality manifests in the observed sample strongly suggests that, from galaxy to galaxy, the spatial distribution of {\OVI}-absorbing gas is similarly organized with respect to the galaxy major axis.  It also suggests this spatial structure is not averaged out when galaxies of various inclinations are included in the sample.

\begin{figure*}[th]
 \begin{minipage}[b]{1.0\textwidth} 
 \centering   
 \includegraphics[width=0.92\textwidth]{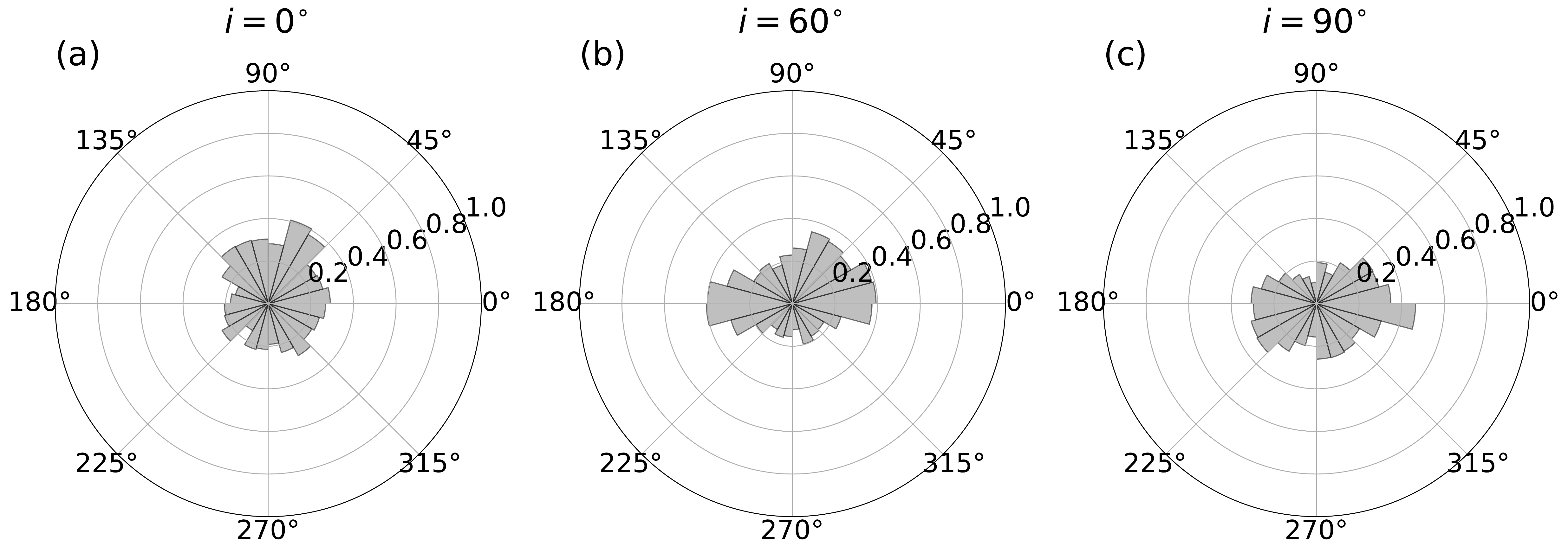}
  \end{minipage}
  \hfill \vglue 0.05in
  \begin{minipage}[b]{1.0\textwidth}
   \centering 
   \includegraphics[width=0.92\textwidth]{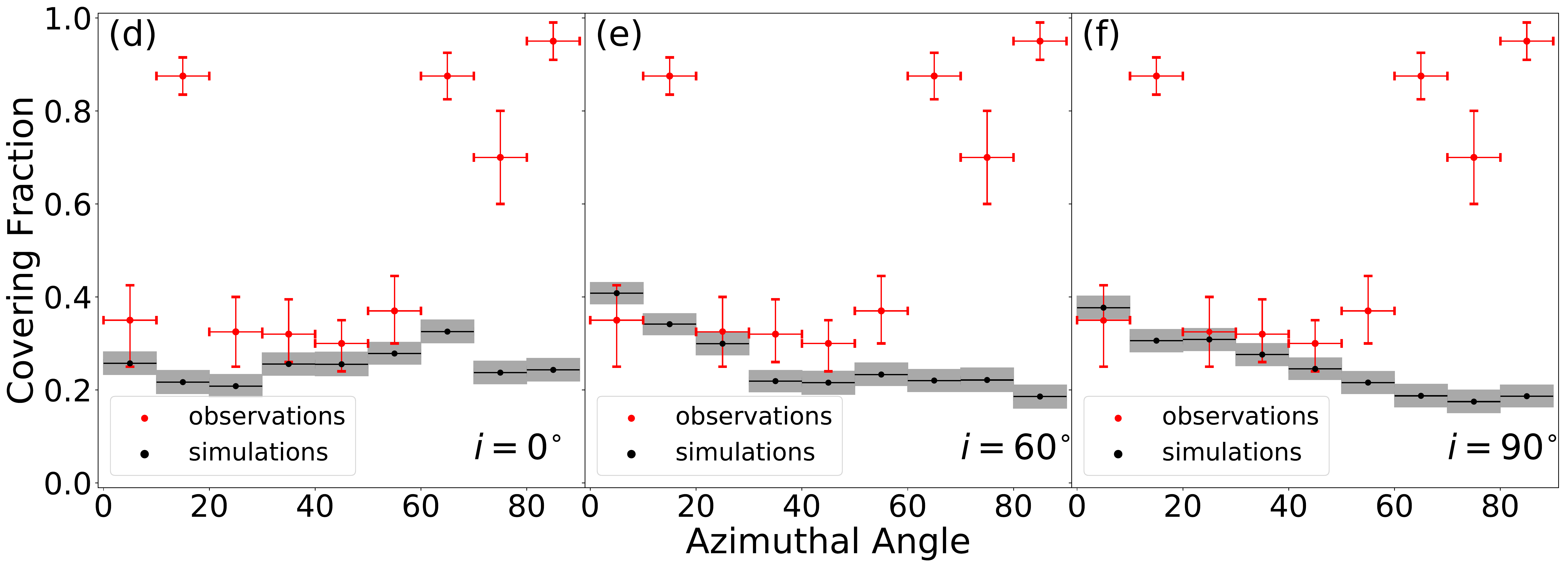}
  \end{minipage}
  \caption{(a,b,c) The sky view of the {\OVI} covering fraction as a function of the full azimuthal angle for the ensemble of simulated galaxies at fixed inclinations $i = 0\degree$,  $60\degree$, and $90\degree$ respectively. (d,e,f) The {\OVI} covering fraction as a function of primary azimuthal angle for the ensemble of simulated galaxies at $i = 0\degree$,  $60\degree$, and $90\degree$ respectively. Each simulated projection contains 1000 sightlines with random $D/R_{vir} < 1.5$ and random azimuthal/position angles. Note for the top panels that the rings indicate covering fraction, with taller bars indicating larger covering fractions.}
  \label{combo}
\end{figure*}

In contrast, our lack of an azimuthal bimodality in the mock sample would suggest that the {\OVI} absorbing gas surrounding simulated galaxies either (1) does not exhibit a common spatial distribution from galaxy to galaxy, or (2) does not exhibit any  structure to its spatial distribution, i.e., is quasi-isotropically distributed. For the former, each simulated galaxy could exhibit a well-defined spatial distribution, but this distribution would not be oriented similarly from galaxy to galaxy with respect to the galactic disk. In this case, ``observing'' the absorption through simulated galaxies of various inclinations might average out the covering fraction as a function of azimuthal angle. 
In the latter case, the amorphous spatial distribution of the absorbing gas would naturally not produce any modes in the covering fraction as a function of azimuthal angle.

In order to improve our understanding of the {\OVI} covering fraction distribution as a function of position angle obtained for the mock sample, we examined the full $360\degree$ azimuthal distribution of the {\OVI} absorption covering fraction for each of the simulated galaxies as a function of inclination angle.  The sky projection of the galaxies was defined so that the axis of rotation for the inclination is in the plane of the sky (perpendicular to the observer). 

In Figure~\ref{grid_plots}, we present the full sky galaxy-to-galaxy azimuthal variation in the covering fraction in azimuthal bins of $15\degree$,  where a line connecting the nodes $0\degree$ and $180\degree$ defines the projected major axes of the galaxies on the plane of the sky (corresponding to the projection of the galaxy disks), and where the nodes $90\degree$ and $270\degree$ define the sky-projected minor axes of the galaxies (corresponding to the projected polar axes of the galaxies.) For each simulated galaxy we present the covering fraction at three sky-projected inclination angles, $i=0\degree$, $60\degree$, and $90\degree$. For each galaxy projection, all 1000 LOS with random distributions of $D/R_{vir} \leq 1.5$ and azimuthal angle, are included.

We see that, around a given galaxy, the {\OVI} absorbing gas is not uniformly distributed, and that it has clear spatial structure as inferred from absorption line analysis. For each galaxy, the covering fraction exhibits substantial variation with azimuthal angle and galaxy inclination. We also find that, for a fixed inclination (or sky viewing angle), the azimuthal distribution of covering fraction varies substantially from galaxy to galaxy. 

It would seem that one outcome of constructing the mock samples is an azimuthal averaging of the covering fraction. Indeed, as presented in Figure~\ref{combo}(a--c), we find that when multiple simulated galaxies are observed at a given inclination angle (i.e., many lines of sight through several galaxies of identical inclination), the azimuthal distribution of the {\OVI} covering fraction becomes significantly more symmetric for each of the three illustrated inclinations.  

Observationally, the full $360\degree$ azimuthal distribution cannot be studied due to degeneracy of the four polar quadrants; only the primary angle ($0\degree \leq \Phi \leq 90\degree$) can be studied. For the $360\degree$ distributions shown in Figure~\ref{combo}(a--c), we present the covering fraction as a function of the primary azimuthal angle in Figure~\ref{combo}(d--f). These are not comparable to the full mock samples shown in Figure~\ref{cf_phi}, as these distributions are for 1000 LOS and fixed galaxy inclinations, whereas the mock sample comprises 15,000 realizations of 53 LOS drawn from the observed distribution of impact parameters and inclinations. The simulated galaxies yield a relatively flat azimuthal distribution of {\OVI} covering fraction for each of the three presented inclination angles ($i=0\degree$, $60\degree$, and $90\degree$), although at both $i=60\degree$ and $90\degree$ there is a slightly higher covering fraction along the galaxy major axis ($ \Phi = 0\degree, 180\degree$). For comparison, we also present the observed {\OVI} covering fractions in Figures~\ref{combo}(d--f) and note that the {\OVI} covering fraction of the simulated galaxies matches observations well in the azimuthal region $20\degree \leq \Phi \leq 60\degree$. As shown in Figure~\ref{cf_D}, there is still a general trend of decreasing {\OVI} covering fraction with increasing azimuthal angle. 

In summary, as inferred from Figure~\ref{grid_plots}, we find that individual simulated galaxies have (1) clearly defined and detectable sky-projected spatial structure in {\OVI} absorption, and (2) unique spatial structure from galaxy to galaxy with no common orientation relative to the galactic disk. As such, the process of adopting ``many lines of sight through several galaxies'' averages out the sky-projection of this spatial structure, even when the inclination of the galaxies is a controlled variable (see Figure \ref{combo}). Given the complicated nature of the CGM, we might expect that observed galaxies would exhibit similar, somewhat randomized structure that would wash out any trends in covering fraction with azimuthal angle.  However, the observed bimodality tantalizingly suggests that ``real-world'' galaxies may have similar spatial structure from galaxy to galaxy.

\subsection{Simulating the Azimuthal Bimodality in the CGM}

From the work of \citet{Kacprzak_2019}, we know that the {\OVI}-absorbing gas surrounding the simulated galaxies in the VELA simulations resides in two main spatial/kinematic features. The first are ouflowing and recycling gas structures within CGM, which reside roughly within $1R_{vir}$ of the galaxies.  The second are thick inflowing filaments that exist at further galactocentric distances, but penetrate into the CGM.  While these structures have been identified in a larger sample of the simulated VELA galaxies than studied here, our analysis indicates that such structures are at different spatial locations for different simulated galaxies. Indeed, our analysis suggests that such structures can give rise to {\OVI} absorption and can be ``observed'' by examining the sky-projected covering fraction of the {\OVI} absorption (as seen in Figure \ref{grid_plots}).

Using EAGLE simulations and TNG50 of the IllustrisTNG simulations, \citet{peroux20} recently showed outflow gas has preferentially higher metallicity than inflowing gas.  They further show higher metallicity gas along the minor axis, suggestive of enrichment of gas from galactic outflows, and lower metallicity accreting gas along the major axis.  This bimodality becomes stronger with increasing impact parameter. 
In the VELA simulations, we do not see an increase of {\OVI} along the minor axis (as is seen in the bimodality in our observed sample), however, we also show (see Figure~\ref{comboOVI}) that {\OVI} outflows do not extend significantly past $D/R_{vir} > 0.375$.  As the majority of the out-flowing gas is not making it out $D/R_{vir} \approx 0.375$ in the VELA simulations, this may potentially explain why the VELA simulations do not reproduce the bimodal covering fraction distribution with azimuthal angle.

One contributing factor as to why the bimodal spatial covering fraction is not seen in the VELA simulations could be that the simulated galaxies are studied at $z\simeq 1$, whereas the mean redshift of the observed galaxies is $z\simeq 0.3$. As redshift decreases the angular area subtended by infalling filamentary streams is expected to increase \citep[e.g.,][]{dekel09, danovich12}, and this could yield a corresponding evolutionary change in the azimuthal distribution of the {\OVI} covering fraction from $z\simeq 1$ to $z\simeq 0.3$. 

A gentler more spread out accretion could yield a greater influence of the galaxy stellar feedback in governing the spatial distribution of the {\OVI}-bearing gas. Potentially, this could cause the spatially stochastic structure at $z\simeq 1$ (see Figure~\ref{grid_plots}) to transform into a systematic structure that gives rise to the observed azimuthal bimodility by $z\simeq 0.3$ \citep[as was seen by][]{peroux20}. While we attempted to control for the influence of mergers by selecting simulated galaxies that had no major-mergers for up to 1~Gyr prior to the redshift of study, it is currently not possible to directly test the VELA simulations below $z=1$ as to whether redshift evolution might resolve the fact that we do not find a bimodal azimuthal covering fraction.

All things considered, we can only speculate how the CGM of the VELA simulations we have studied would evolve from $z\simeq 1$ to $\simeq 0.3$. Observational data and theory are somewhat contradictory on guiding our insights. Though higher {\OVI} column density appears to be associated with star forming galaxies with specific-SFR greater than $10^{-11}$~yr$^{-1}$ \citep{tumlinson11}, it is hard to reconcile this finding with growing observation evidence \citep{bielby19, Zahedy19, NG2019} and the theoretical expectation \citep{Oppenheimer16, Nelson2018, oppenheimer18, Qu18} that the {\OVI} column density peaks for galaxies with halo mass $\log M_h/M_{\odot} \simeq 12$.   

It is difficult to interpret the relationship between {\OVI} absorbing CGM gas and stellar driven outflows and accretion in view of the observed gas kinematics and metallicities.  \citet{nielsen17} find that the velocity dispersion of the gas has no systematic differences with galaxy orientation or color, \citet{Kacprzak_2019} find that the global kinematics of the gas shows no coupling to the host galaxy rotation, and \citet{Pointon_2019} find that there is no clear metallicity trend between minor axis and major axis gas.

Despite the difference in redshift of the VELA simulations and the observed galaxy sample, since galaxies mostly undergo passive evolution from redshift $z\simeq1$ to $z\simeq0.3$ \citep[e.g.,][]{satoh19}, and since the cooling time of the hot, diffuse gas is longer than a Hubble time \citep{sutherland93, wiersma09}, we do not expect that our findings would dramatically change. While there is a dependency of {\OVI} absorption on halo mass \citep{NG2019} the mean virial mass of the observed and simulated galaxies are quite similar ($\log(M_{vir}/M_{\odot}) = 11.8, 11.6$ respectively) so we expect this to not significantly effect our results.

One of the greatest recent challenges is to duplicate the azimuthal spatial dependence, as discussed throughout this work. Interestingly, it is only very recently that a subset of simulations have achieved outflows concentrated along the galaxy minor axis \citep{nelson19, mitchell20, peroux20}.  However, it remains to be seen exactly how these spatial-kinematic CGM structures translate to covering fractions obtained through mock absorption line experiments (using many LOS through many simulated galaxies).


It is only recently that pioneering studies by \citet{Hummels19, peeples19, vandevoort19} have investigated resolution effects on the nature of the simulated CGM gas. \citet{peeples19} finds that absorption lines arise in gas structures with masses below $10^4$~M$_{\odot}$, while \citet{Hummels19} predicts CGM gas structure with sub-~M$_{\odot}$, both substantially lower than mass resolution of zoom simulations, such as the VELA simulations. 

Improvements of the VELA simulations used for this work, which modeled only thermal-energy supernova feedback and radiative feedback from stars, include adding a non-thermal radiation pressure to the total gas pressure in regions where ionizing photons from massive stars are produced.  Modern VELA simulations have been run with enhanced momentum as part of the supernova feedback \citep[see][]{gentry17}, and this new feedback mechanism yields an improved stellar mass-halo mass relations (Ceverino {\etal}~2020, in prep).  It remains to be seen whether the new feedback mechanisms generate a more realistic CGM relative to the VELA simulations used in this work. We will compare the results of this paper to this new generation of the VELA simulations in future work.

\section{Conclusions}
\label{Conclusions}

For this study we designed an experiment to emulate the Multiphase Galaxy Halos Survey \citepalias[][comprising 53 galaxies]{kacprzak15} using four Milky Way-like galaxies from the VELA hydrodynamic cosmological simulations of \citet{ceverino14}. Our primary goal was to examine whether the observed impact parameter and the bimodal azimuthal distributions of {\OVI} covering fractions (reproduced in Figures~\ref{cf_D} and \ref{cf_phi}, respectively) could be ``observed'' for the simulated galaxies. In other words, we investigated whether insights into the observed spatial distribution of {\OVI} absorbing gas in the CGM could be gained by ``observing'' these distributions in simulations.

We used the {\sc Mockspec} quasar absorption line analysis pipeline \citep{churchill15, rachel_thesis} to generate LOS through the CGM of the VELA galaxies, generate absorption spectra, detect and measure the absorption line properties, and identify the gas cells in the simulations that contribute to the detected absorption (i.e., the absorbing cells). To make a fair comparison to the observed galaxy-absorber sample, we constructed and studied a mock sample. For this mock sample, we calculated the mean covering fractions and their $1\sigma$ confidence levels using 15,000 realizations of a sample of 53 LOS through the simulated galaxies.  For each realization, the distributions of sky-projected LOS impact parameters, galaxy inclinations, and LOS-galaxy azimuthal angles are generated to be statistically consistent with the observed sample of galaxies.  

In summary, for our comparison of the observed and the mock samples, we found 
\begin{enumerate}
\setcounter{enumi}{0}

\item The {\OVI} absorption covering fraction as a function of impact parameter normalized by the virial radius is statistically consistent with observations in the inner CGM ($D/R_{vir} \leq 0.375$), but in the intermediate range ($0.375 < D/R_{vir} \leq 0.75$), the mock sample yielded too low of a covering fraction by a factor of three to four.  In the outer CGM ($D/R_{vir} > 0.75$), the covering fraction was consistent when taking uncertainties into account, as the values fell within $1.2\sigma$ of the observed values. 

\item The {\OVI} covering fraction in the simulations as a function of azimuthal angle had no modes, but is a smooth flat distribution with slightly decreasing value from $\Phi =0\degree$ (projected major axis) to $\Phi =90\degree$ (projected minor axis), in contrast to the bimodal distribution seen in the observations.  For intermediate azimuthal angles ($20\degree \leq \Phi \leq 60\degree$) the mean simulated covering fraction was consistent with the observed values.  

\end{enumerate}

We undertook additional investigations designed to characterise the gas phase, chemical, ionization, spatial, and kinematic distributions of {\OVI}-absorbing gas associated with simulated galaxies as inferred from the absorption line analysis techniques. Part of our motivation was to (1) investigate possible reasons for the relative paucity of {\OVI} absorption in the simulated CGM at $0.375 < D/R_{vir} \leq 0.75$, and (2) explore whether our experimental method of creating a mock sample using the assumption that ``many lines of sight through a few galaxies is equivalent to one line of sight through many galaxies.''  To this end, we studied the average {\OVI} mass surface density of absorbing gas and the mass distribution of {\OVI} for the total radial velocity and the radial velocity unit vector for {\OVI}-absorbing gas. We also examined the azimuthal {\OVI} covering fractions for the individual simulated galaxies at various fixed inclination angles and compared these to the ``averaged'' azimuthal covering fraction distribution for the ensemble of simulated galaxies.  Our findings are summarized as follows:

\begin{enumerate}
\setcounter{enumi}{2}

\item  The CGM gas selected by {\OVI} absorption in the VELA simulations has gas phase conditions  $4 \leq \log T/$K$~\leq 6.5$ and $-5 \leq \log n_{\tH}/{\rm cm}^{-3}  \leq -1$ and ionization fractions on the order of $\log f_{\tOVI} \simeq 0.1$. The majority of this {\OVI}-absorbing gas mass resides within $4 \leq \log T/$K$~\leq 4.5$ and $-4.5 \leq \log n_{\tH}/{\rm cm}^{-3}\leq -3.5$ and is predominantly photoionized. The {\OVI}-absorbing gas comprises the majority of the gas mass in the CGM (about 80\%), but only a minority of the CGM volume (about 20\%) with inside the virial radius.  In stark contrast to the {\OVI}-absorbing gas, which has a peak metallicity at $\log Z/Z_{\odot} = -0.7$, the highest metallicity ($Z \simeq 1.2 Z_{\odot}$) component of the CGM is the hot, diffuse, non-absorbing gas.

\item We found that the {\OVI} mass surface density of gas selected by {\OVI} absorption, $\bar{\Sigma} ({\OVI}$-absorbing gas), shows a steady and smooth decline from the inner to the outer CGM, but the outflowing gas shows a dramatic factor of four drop from the inner to the outer CGM at $D/R_{vir} \simeq 0.375$. We infer that the majority of the outflowing gas does not reach the outer regions of the CGM (only $\sim\! 20$\% is reaching to $D/R_{vir} > 0.375$) and this explains the decrease in {\OVI} covering fraction for $D/R_{vir} > 0.375$ in the mock sample.

\item  The gas kinematics indicate that, for $D/R_{vir} \leq 0.375$, roughly 70\% of the total {\OVI} mass is outflow, and roughly $40$\% of the total {\OVI} mass is an outflow on a radial trajectory with radial velocities between $120 \leq v_r \leq 230$~{\kms}.  For $0.375 < D/R_{vir} \simeq 0.75$, there is only about one-fourth as much {\OVI} mass, of which roughly half is outflowing. Corroborating the behavior of the {\OVI} mass surface density with increasing $D/R_{vir}$, the kinematics suggest that the drop in {\OVI}-absorbing gas mass with $D/R_{vir}$ is because the majority of the outflowing gas is not making it out past $D/R_{vir}\simeq 0.375$, resulting in the drop in the {\OVI} covering fraction in the mock sample.  

\item As inferred from absorption line analysis, a given simulated galaxy has a unique spatial structure of {\OVI} absorbing gas.  For a given galaxy, the azimuthal distribution of the covering fraction changes as a function of inclination, and for a fixed inclination the azimuthal distribution of covering fraction varies substantially from galaxy to galaxy. When an ensemble of simulated galaxies is ``observed'', these individual structures are averaged out, resulting in an azimuthal distribution of covering fraction that has no modes. Our experiment shows that the azimuthal asymmetries in the individual simulated galaxies do not reflect the universal bimodality observed for real world galaxies. 
\end{enumerate}

In the future, we plan to study the next generation of VELA simulations to examine {\OVI} and a broader range of ions. We also plan on examining the kinematics of the simulations, using methods of \citet{nielsen17}, i.e., the pixel-velocity two-point correlation function (pixel-velocity TPCF), to study the ability of the simulations to match observations of the gas dynamics.

We would suggest that running the VELA simulations to $z \simeq 0$ would be highly informative for comparing the properties of the CGM resulting from different feedback recipes being implemented in the most current VELA simulations as well as future versions \citep[see][and Ceverino {\etal}~2020, in prep]{gentry17}.  This is especially important as it will allow us to compare the simulations directly with {\it HST\/} UV absorption line spectra of {\HI}, {\CIV}, and {\OVI} transitions, which we plan to do in future work.

\begin{acknowledgements}
We thank the referee for helpful comments that improved this manuscript. This material is based upon work supported by the National Science Foundation under Grant No.\ 1517816 issued to CWC and JCC. GGK and NMN acknowledge the support of the Australian Research Council through a Discovery Project DP170103470. Parts of this research were supported by the Australian Research Council Centre of Excellence for All Sky Astrophysics in 3 Dimensions (ASTRO 3D), through project number CE170100013. DC is a Ramon-Cajal fellow. SM is supported by the Humboldt Foundation (Germany) through the Experienced Researchers Fellowship.

\end{acknowledgements}

\bibliographystyle{apj}

\begin{thebibliography}{102}
\expandafter\ifx\csname natexlab\endcsname\relax\def\natexlab#1{#1}\fi

\bibitem[{{Agertz} \& {Kravtsov}(2015)}]{Agertz15}
{Agertz}, O., \& {Kravtsov}, A.~V. 2015, \apj, 804, 18

\bibitem[{{Asplund} {et~al.}(2009){Asplund}, {Grevesse}, {Sauval}, \&
  {Scott}}]{Asplund09}
{Asplund}, M., {Grevesse}, N., {Sauval}, A.~J., \& {Scott}, P. 2009, \araa, 47,
  481

\bibitem[{{Behroozi} {et~al.}(2010){Behroozi}, {Conroy}, \&
  {Wechsler}}]{Behroozi10}
{Behroozi}, P.~S., {Conroy}, C., \& {Wechsler}, R.~H. 2010, \apj, 717, 379

\bibitem[{{Behroozi} {et~al.}(2013){Behroozi}, {Wechsler}, \&
  {Conroy}}]{behroozi13}
{Behroozi}, P.~S., {Wechsler}, R.~H., \& {Conroy}, C. 2013, \apj, 770, 57

\bibitem[{{Bergeron} \& {Herbert-Fort}(2005)}]{bergeron05}
{Bergeron}, J., \& {Herbert-Fort}, S. 2005, ArXiv Astrophysics e-prints

\bibitem[{{Bielby} {et~al.}(2019){Bielby}, {Stott}, {Cullen}, {Tripp},
  {Burchett}, {Fumagalli}, {Morris}, {Tejos}, {Crain}, {Bower}, \&
  {Prochaska}}]{bielby19}
{Bielby}, R.~M., {Stott}, J.~P., {Cullen}, F., {et~al.} 2019, \mnras, 486, 21

\bibitem[{{Birnboim} \& {Dekel}(2003)}]{birnboim03}
{Birnboim}, Y., \& {Dekel}, A. 2003, \mnras, 345, 349

\bibitem[{{Bond} {et~al.}(1996){Bond}, {Kofman}, \& {Pogosyan}}]{Bond96}
{Bond}, J.~R., {Kofman}, L., \& {Pogosyan}, D. 1996, \nat, 380, 603

\bibitem[{{Bordoloi} {et~al.}(2014){Bordoloi}, {Tumlinson}, {Werk},
  {Oppenheimer}, {Peeples}, {Prochaska}, {Tripp}, {Katz}, {Dav{\'e}}, {Fox},
  {Thom}, {Ford}, {Weinberg}, {Burchett}, \& {Kollmeier}}]{bordoloi-cosdwarfs}
{Bordoloi}, R., {Tumlinson}, J., {Werk}, J.~K., {et~al.} 2014, \apj, 796, 136

\bibitem[{Bothwell {et~al.}(2013)Bothwell, Maiolino, Kennicutt, Cresci,
  Mannucci, Marconi, \& Cicone}]{Bothwell13}
Bothwell, M.~S., Maiolino, R., Kennicutt, R., J., {et~al.} 2013, Monthly
  Notices of the Royal Astronomical Society, 433, 1425

\bibitem[{{Bouch{\'e}} {et~al.}(2012){Bouch{\'e}}, {Hohensee}, {Vargas},
  {Kacprzak}, {Martin}, {Cooke}, \& {Churchill}}]{bouche12}
{Bouch{\'e}}, N., {Hohensee}, W., {Vargas}, R., {et~al.} 2012, \mnras, 426, 801

\bibitem[{{Bryan} \& {Norman}(1998)}]{bryan98}
{Bryan}, G.~L., \& {Norman}, M.~L. 1998, \apj, 495, 80

\bibitem[{{Bundy} {et~al.}(2005){Bundy}, {Ellis}, \& {Conselice}}]{Bundy05}
{Bundy}, K., {Ellis}, R.~S., \& {Conselice}, C.~J. 2005, \apj, 625, 621

\bibitem[{{Ceverino} {et~al.}(2016){Ceverino}, {Arribas}, {Colina},
  {Rodr{\'{\i}}guez Del Pino}, {Dekel}, \& {Primack}}]{Ceverino16}
{Ceverino}, D., {Arribas}, S., {Colina}, L., {et~al.} 2016, \mnras, 460, 2731

\bibitem[{{Ceverino} {et~al.}(2010){Ceverino}, {Dekel}, \&
  {Bournaud}}]{Ceverino10}
{Ceverino}, D., {Dekel}, A., \& {Bournaud}, F. 2010, \mnras, 404, 2151

\bibitem[{{Ceverino} \& {Klypin}(2009)}]{ceverino09}
{Ceverino}, D., \& {Klypin}, A. 2009, \apj, 695, 292

\bibitem[{{Ceverino} {et~al.}(2014){Ceverino}, {Klypin}, {Klimek},
  {Trujillo-Gomez}, {Churchill}, {Primack}, \& {Dekel}}]{ceverino14}
{Ceverino}, D., {Klypin}, A., {Klimek}, E.~S., {et~al.} 2014, \mnras, 442, 1545

\bibitem[{Chen(2012)}]{chen12}
Chen, H.-W. 2012, Monthly Notices of the Royal Astronomical Society, 427, 1238

\bibitem[{{Chen} {et~al.}(2010){Chen}, {Helsby}, {Gauthier}, {Shectman},
  {Thompson}, \& {Tinker}}]{chen10a}
{Chen}, H.-W., {Helsby}, J.~E., {Gauthier}, J.-R., {et~al.} 2010, \apj, 714,
  1521

\bibitem[{{Chen} {et~al.}(2001){Chen}, {Lanzetta}, \& {Webb}}]{chen01a}
{Chen}, H.-W., {Lanzetta}, K.~M., \& {Webb}, J.~K. 2001, \apj, 556, 158

\bibitem[{{Churchill} {et~al.}(2012){Churchill}, {Kacprzak}, {Steidel},
  {Spitler}, {Holtzman}, {Nielsen}, \& {Trujillo-Gomez}}]{cwc1317b}
{Churchill}, C.~W., {Kacprzak}, G.~G., {Steidel}, C.~C., {et~al.} 2012, \apj,
  760, 68

\bibitem[{{Churchill} {et~al.}(2014){Churchill}, {Klimek}, {Medina}, \& {Vander
  Vliet}}]{cwc14}
{Churchill}, C.~W., {Klimek}, E., {Medina}, A., \& {Vander Vliet}, J.~R. 2014,
  ArXiv e-prints

\bibitem[{{Churchill} {et~al.}(2000){Churchill}, {Mellon}, {Charlton},
  {Jannuzi}, {Kirhakos}, {Steidel}, \& {Schneider}}]{churchill00}
{Churchill}, C.~W., {Mellon}, R.~R., {Charlton}, J.~C., {et~al.} 2000, \apj,
  543, 577

\bibitem[{{Churchill} {et~al.}(2013){Churchill}, {Nielsen}, {Kacprzak}, \&
  {Trujillo-Gomez}}]{churchill13}
{Churchill}, C.~W., {Nielsen}, N.~M., {Kacprzak}, G.~G., \& {Trujillo-Gomez},
  S. 2013, \apjl, 763, L42

\bibitem[{{Churchill} {et~al.}(2015){Churchill}, {Vander Vliet},
  {Trujillo-Gomez}, {Kacprzak}, \& {Klypin}}]{churchill15}
{Churchill}, C.~W., {Vander Vliet}, J.~R., {Trujillo-Gomez}, S., {Kacprzak},
  G.~G., \& {Klypin}, A. 2015, \apj, 802, 10

\bibitem[{{Danovich} {et~al.}(2012){Danovich}, {Dekel}, {Hahn}, \&
  {Teyssier}}]{danovich12}
{Danovich}, M., {Dekel}, A., {Hahn}, O., \& {Teyssier}, R. 2012, \mnras, 422,
  1732

\bibitem[{{Dav{\'e}} {et~al.}(2011){Dav{\'e}}, {Oppenheimer}, \&
  {Finlator}}]{dave11a}
{Dav{\'e}}, R., {Oppenheimer}, B.~D., \& {Finlator}, K. 2011, \mnras, 415, 11

\bibitem[{{Dekel} \& {Birnboim}(2006)}]{Dekel06}
{Dekel}, A., \& {Birnboim}, Y. 2006, \mnras, 368, 2

\bibitem[{{Dekel} {et~al.}(2009{\natexlab{a}}){Dekel}, {Sari}, \&
  {Ceverino}}]{Dekel09apj}
{Dekel}, A., {Sari}, R., \& {Ceverino}, D. 2009{\natexlab{a}}, \apj, 703, 785

\bibitem[{{Dekel} {et~al.}(2009{\natexlab{b}}){Dekel}, {Birnboim}, {Engel},
  {Freundlich}, {Goerdt}, {Mumcuoglu}, {Neistein}, {Pichon}, {Teyssier}, \&
  {Zinger}}]{dekel09}
{Dekel}, A., {Birnboim}, Y., {Engel}, G., {et~al.} 2009{\natexlab{b}}, \nat,
  457, 451

\bibitem[{Doroshkevich {et~al.}(1980)Doroshkevich, Kotok, Novikov, Polyudov,
  Shandarin, \& Sigov}]{Doroshkevich80}
Doroshkevich, A.~G., Kotok, E.~V., Novikov, I.~D., {et~al.} 1980, Monthly
  Notices of the Royal Astronomical Society, 192, 321

\bibitem[{{Draine}(2011)}]{Draine}
{Draine}, B.~T. 2011, Physics of the Interstellar and Intergalactic Medium
  (Princeton University Press)

\bibitem[{{Fall} \& {Efstathiou}(1980)}]{Fall80}
{Fall}, S.~M., \& {Efstathiou}, G. 1980, \mnras, 193, 189

\bibitem[{{Ferland} {et~al.}(1998){Ferland}, {Korista}, {Verner}, {Ferguson},
  {Kingdon}, \& {Verner}}]{Ferland98}
{Ferland}, G.~J., {Korista}, K.~T., {Verner}, D.~A., {et~al.} 1998, \pasp, 110,
  761

\bibitem[{{Ferland} {et~al.}(2013){Ferland}, {Porter}, {van Hoof}, {Williams},
  {Abel}, {Lykins}, {Shaw}, {Henney}, \& {Stancil}}]{Ferland13}
{Ferland}, G.~J., {Porter}, R.~L., {van Hoof}, P.~A.~M., {et~al.} 2013, Revista
  Mexicana de Astronomia y Astrofisica, 49, 137

\bibitem[{{Ford} {et~al.}(2013){Ford}, {Oppenheimer}, {Dav{\'e}}, {Katz},
  {Kollmeier}, \& {Weinberg}}]{ford13}
{Ford}, A.~B., {Oppenheimer}, B.~D., {Dav{\'e}}, R., {et~al.} 2013, \mnras,
  432, 89

\bibitem[{{Gentry} {et~al.}(2017){Gentry}, {Krumholz}, {Dekel}, \&
  {Madau}}]{gentry17}
{Gentry}, E.~S., {Krumholz}, M.~R., {Dekel}, A., \& {Madau}, P. 2017, \mnras,
  465, 2471

\bibitem[{{Haardt} \& {Madau}(2005)}]{HaardtMadau2005}
{Haardt}, F., \& {Madau}, P. 2005, unpublished spectra in 2005 August update to
  Haardt \& Madau (2001) and included in the photoionization code CLOUDY

\bibitem[{{Hafen} {et~al.}(2020){Hafen}, {Faucher-Gigu{\`e}re},
  {Angl{\'e}s-Alc{\'a}zar}, {Stern}, {Kere{\v{s}}}, {Esmerian}, {Wetzel},
  {El-Badry}, {Chan}, \& {Murray}}]{hafen20}
{Hafen}, Z., {Faucher-Gigu{\`e}re}, C.-A., {Angl{\'e}s-Alc{\'a}zar}, D.,
  {et~al.} 2020, \mnras, 494, 3581

\bibitem[{{Hummels} {et~al.}(2019){Hummels}, {Smith}, {Hopkins}, {O'Shea},
  {Silvia}, {Werk}, {Lehner}, {Wise}, {Collins}, \& {Butsky}}]{Hummels19}
{Hummels}, C.~B., {Smith}, B.~D., {Hopkins}, P.~F., {et~al.} 2019, \apj, 882,
  156

\bibitem[{{Kacprzak}(2017)}]{kacprzak17}
{Kacprzak}, G.~G. 2017, Astrophysics and Space Science Library, Vol. 430, {Gas
  Accretion in Star-Forming Galaxies}, ed. A.~{Fox} \& R.~{Dav{\'e}}, 145

\bibitem[{{Kacprzak} {et~al.}(2010){Kacprzak}, {Churchill}, {Ceverino},
  {Steidel}, {Klypin}, \& {Murphy}}]{ggk-sims}
{Kacprzak}, G.~G., {Churchill}, C.~W., {Ceverino}, D., {et~al.} 2010, \apj,
  711, 533

\bibitem[{{Kacprzak} {et~al.}(2012{\natexlab{a}}){Kacprzak}, {Churchill}, \&
  {Nielsen}}]{kcn12}
{Kacprzak}, G.~G., {Churchill}, C.~W., \& {Nielsen}, N.~M. 2012{\natexlab{a}},
  \apjl, 760, L7

\bibitem[{{Kacprzak} {et~al.}(2012{\natexlab{b}}){Kacprzak}, {Churchill}, \&
  {Nielsen}}]{kacprzak12}
---. 2012{\natexlab{b}}, \apjl, 760, L7

\bibitem[{{Kacprzak} {et~al.}(2008){Kacprzak}, {Churchill}, {Steidel}, \&
  {Murphy}}]{Kacprzak08}
{Kacprzak}, G.~G., {Churchill}, C.~W., {Steidel}, C.~C., \& {Murphy}, M.~T.
  2008, \aj, 135, 922

\bibitem[{{Kacprzak} {et~al.}(2015){Kacprzak}, {Muzahid}, {Churchill},
  {Nielsen}, \& {Charlton}}]{kacprzak15}
{Kacprzak}, G.~G., {Muzahid}, S., {Churchill}, C.~W., {Nielsen}, N.~M., \&
  {Charlton}, J.~C. 2015, \apj, 815, 22

\bibitem[{Kacprzak {et~al.}(2019)Kacprzak, Vliet, Nielsen, Muzahid, Pointon,
  Churchill, Ceverino, Arraki, Klypin, Charlton, \& Lewis}]{Kacprzak_2019}
Kacprzak, G.~G., Vliet, J. R.~V., Nielsen, N.~M., {et~al.} 2019, The
  Astrophysical Journal, 870, 137

\bibitem[{{Katz} {et~al.}(2003){Katz}, {Keres}, {Dave}, \& {Weinberg}}]{Katz03}
{Katz}, N., {Keres}, D., {Dave}, R., \& {Weinberg}, D.~H. 2003, Astrophysics
  and Space Science Library, Vol. 281, {How Do Galaxies Get Their Gas?}, ed.
  J.~L. {Rosenberg} \& M.~E. {Putman}, 185

\bibitem[{{Kere{\v s}} {et~al.}(2009){Kere{\v s}}, {Katz}, {Fardal},
  {Dav{\'e}}, \& {Weinberg}}]{keres09}
{Kere{\v s}}, D., {Katz}, N., {Fardal}, M., {Dav{\'e}}, R., \& {Weinberg},
  D.~H. 2009, \mnras, 395, 160

\bibitem[{{Kere{\v s}} {et~al.}(2005){Kere{\v s}}, {Katz}, {Weinberg}, \&
  {Dav{\'e}}}]{keres05}
{Kere{\v s}}, D., {Katz}, N., {Weinberg}, D.~H., \& {Dav{\'e}}, R. 2005,
  \mnras, 363, 2

\bibitem[{{Klypin} {et~al.}(2001){Klypin}, {Kravtsov}, {Bullock}, {Primack}, \&
  {Klypin}}]{Klypin2001}
{Klypin}, A., {Kravtsov}, A.~V., {Bullock}, J., {Primack}, C.~C., \& {Klypin},
  J. 2001, \apj, 554, 903

\bibitem[{{Klypin} \& {Shandarin}(1983)}]{Klypin83}
{Klypin}, A.~A., \& {Shandarin}, S.~F. 1983, \mnras, 204, 891

\bibitem[{{Kravstov}(1999)}]{Kravstov99_thesis}
{Kravstov}, A.~V. 1999, PhD thesis, New Mexico State University

\bibitem[{Kravtsov {et~al.}(2018)Kravtsov, Vikhlinin, \&
  Meshscheryakov}]{Kravtsov14}
Kravtsov, A., Vikhlinin, A., \& Meshscheryakov, A. 2018, Astron. Lett., 44, 8

\bibitem[{Kravtsov(2003)}]{Kravtsov_2003}
Kravtsov, A.~V. 2003, The Astrophysical Journal, 590, L1

\bibitem[{{Kravtsov} {et~al.}(1997){Kravtsov}, {Klypin}, \&
  {Khokhlov}}]{Kravstov97}
{Kravtsov}, A.~V., {Klypin}, A.~A., \& {Khokhlov}, A.~M. 1997, \apjs, 111, 73

\bibitem[{{Lilly} {et~al.}(2013){Lilly}, {Carollo}, {Pipino}, {Renzini}, \&
  {Peng}}]{lilly-bathtub}
{Lilly}, S.~J., {Carollo}, C.~M., {Pipino}, A., {Renzini}, A., \& {Peng}, Y.
  2013, \apj, 772, 119

\bibitem[{{Lodders}(2019)}]{Lodder19}
{Lodders}, K. 2019, arXiv e-prints, arXiv:1912.00844

\bibitem[{Mannucci {et~al.}(2010)Mannucci, Cresci, Maiolino, Marconi, \&
  Gnerucci}]{Mannucci10}
Mannucci, F., Cresci, G., Maiolino, R., Marconi, A., \& Gnerucci, A. 2010,
  Monthly Notices of the Royal Astronomical Society, 408, 2115

\bibitem[{{Mitchell} {et~al.}(2020){Mitchell}, {Schaye}, \&
  {Bower}}]{mitchell20}
{Mitchell}, P.~D., {Schaye}, J., \& {Bower}, R.~G. 2020, \mnras, 497, 4495

\bibitem[{{Mo} {et~al.}(1998){Mo}, {Mao}, \& {White}}]{Mo98}
{Mo}, H.~J., {Mao}, S., \& {White}, S. D.~M. 1998, \mnras, 295, 319

\bibitem[{{Moster} {et~al.}(2013){Moster}, {Naab}, \& {White}}]{moster13}
{Moster}, B.~P., {Naab}, T., \& {White}, S. D.~M. 2013, \mnras, 428, 3121

\bibitem[{{Munshi} {et~al.}(2013){Munshi}, {Governato}, {Brooks},
  {Christensen}, {Shen}, {Loebman}, {Moster}, {Quinn}, \& {Wadsley}}]{munshi13}
{Munshi}, F., {Governato}, F., {Brooks}, A.~M., {et~al.} 2013, \apj, 766, 56

\bibitem[{Nelson {et~al.}(2018)Nelson, Kauffmann, Pillepich, Genel, Springel,
  Pakmor, Hernquist, Weinberger, Torrey, Vogelsberger, \&
  Marinacci}]{Nelson2018}
Nelson, D., Kauffmann, G., Pillepich, A., {et~al.} 2018, Monthly Notices of the
  Royal Astronomical Society, 477, 450

\bibitem[{{Nelson} {et~al.}(2019){Nelson}, {Pillepich}, {Springel}, {Pakmor},
  {Weinberger}, {Genel}, {Torrey}, {Vogelsberger}, {Marinacci}, \&
  {Hernquist}}]{nelson19}
{Nelson}, D., {Pillepich}, A., {Springel}, V., {et~al.} 2019, \mnras, 490, 3234

\bibitem[{{Ng} {et~al.}(2019){Ng}, {Nielsen}, {Kacprzak}, {Pointon}, {Muzahid},
  {Churchill}, \& {Charlton}}]{NG2019}
{Ng}, M., {Nielsen}, N.~M., {Kacprzak}, G.~G., {et~al.} 2019, \apj, 886, 66

\bibitem[{{Nielsen} {et~al.}(2013{\natexlab{a}}){Nielsen}, {Churchill}, \&
  {Kacprzak}}]{magiicat2}
{Nielsen}, N.~M., {Churchill}, C.~W., \& {Kacprzak}, G.~G. 2013{\natexlab{a}},
  \apj, 776, 115

\bibitem[{{Nielsen} {et~al.}(2013{\natexlab{b}}){Nielsen}, {Churchill},
  {Kacprzak}, \& {Murphy}}]{magiicat1}
{Nielsen}, N.~M., {Churchill}, C.~W., {Kacprzak}, G.~G., \& {Murphy}, M.~T.
  2013{\natexlab{b}}, \apj, 776, 114

\bibitem[{{Nielsen} {et~al.}(2017){Nielsen}, {Kacprzak}, {Muzahid},
  {Churchill}, {Murphy}, \& {Charlton}}]{nielsen17}
{Nielsen}, N.~M., {Kacprzak}, G.~G., {Muzahid}, S., {et~al.} 2017, \apj, 834,
  148

\bibitem[{{Oppenheimer} \& {Dav{\'e}}(2008)}]{oppenheimer08}
{Oppenheimer}, B.~D., \& {Dav{\'e}}, R. 2008, \mnras, 387, 577

\bibitem[{{Oppenheimer} {et~al.}(2018){Oppenheimer}, {Segers}, {Schaye},
  {Richings}, \& {Crain}}]{oppenheimer18}
{Oppenheimer}, B.~D., {Segers}, M., {Schaye}, J., {Richings}, A.~J., \&
  {Crain}, R.~A. 2018, \mnras, 474, 4740

\bibitem[{{Oppenheimer} {et~al.}(2016){Oppenheimer}, {Crain}, {Schaye},
  {Rahmati}, {Richings}, {Trayford}, {Tumlinson}, {Bower}, {Schaller}, \&
  {Theuns}}]{Oppenheimer16}
{Oppenheimer}, B.~D., {Crain}, R.~A., {Schaye}, J., {et~al.} 2016, \mnras, 460,
  2157

\bibitem[{{Pauls} \& {Melott}(1995)}]{Pauls95}
{Pauls}, J.~L., \& {Melott}, A.~L. 1995, \mnras, 274, 99

\bibitem[{{Peeples} {et~al.}(2019){Peeples}, {Corlies}, {Tumlinson}, {O'Shea},
  {Lehner}, {O'Meara}, {Howk}, {Earl}, {Smith}, {Wise}, \&
  {Hummels}}]{peeples19}
{Peeples}, M.~S., {Corlies}, L., {Tumlinson}, J., {et~al.} 2019, \apj, 873, 129

\bibitem[{{P{\'e}roux} {et~al.}(2020){P{\'e}roux}, {Nelson}, {van de Voort},
  {Pillepich}, {Marinacci}, {Vogelsberger}, \& {Hernquist}}]{peroux20}
{P{\'e}roux}, C., {Nelson}, D., {van de Voort}, F., {et~al.} 2020, \mnras

\bibitem[{Pointon {et~al.}(2019)Pointon, Kacprzak, Nielsen, Muzahid, Murphy,
  Churchill, \& Charlton}]{Pointon_2019}
Pointon, S.~K., Kacprzak, G.~G., Nielsen, N.~M., {et~al.} 2019, The
  Astrophysical Journal, 883, 78

\bibitem[{{Qu} \& {Bregman}(2018)}]{Qu18}
{Qu}, Z., \& {Bregman}, J.~N. 2018, \apj, 862, 23

\bibitem[{{Roca-F{\`a}brega} {et~al.}(2019){Roca-F{\`a}brega}, {Dekel},
  {Faerman}, {Gnat}, {Strawn}, {Ceverino}, {Primack}, {Macci{\`o}}, {Dutton},
  {Prochaska}, \& {Stern}}]{roca19}
{Roca-F{\`a}brega}, S., {Dekel}, A., {Faerman}, Y., {et~al.} 2019, \mnras, 484,
  3625

\bibitem[{{Sathyaprakash} {et~al.}(1996){Sathyaprakash}, {Sahni}, \&
  {Shandarin}}]{Sathyaprakash96}
{Sathyaprakash}, B.~S., {Sahni}, V., \& {Shandarin}, S.~F. 1996, \apjl, 462, L5

\bibitem[{{Satoh} {et~al.}(2019){Satoh}, {Kajisawa}, \& {Himoto}}]{satoh19}
{Satoh}, Y.~K., {Kajisawa}, M., \& {Himoto}, K.~G. 2019, \apj, 885, 81

\bibitem[{{Schneider} {et~al.}(1993){Schneider}, {Hartig}, {Jannuzi},
  {Kirhakos}, {Saxe}, {Weymann}, {Bahcall}, {Bergeron}, {Boksenberg},
  {Sargent}, {Savage}, {Turnshek}, \& {Wolfe}}]{schneider93}
{Schneider}, D.~P., {Hartig}, G.~F., {Jannuzi}, B.~T., {et~al.} 1993, \apjs,
  87, 45

\bibitem[{{Shapiro} {et~al.}(1983){Shapiro}, {Struck-Marcell}, \&
  {Melott}}]{Shapiro83}
{Shapiro}, P.~R., {Struck-Marcell}, C., \& {Melott}, A.~L. 1983, \apj, 275, 413

\bibitem[{{Simard} {et~al.}(2002){Simard}, {Willmer}, {Vogt}, {Sarajedini},
  {Phillips}, {Weiner}, {Koo}, {Im}, {Illingworth}, \& {Faber}}]{simard02}
{Simard}, L., {Willmer}, C.~N.~A., {Vogt}, N.~P., {et~al.} 2002, \apjs, 142, 1

\bibitem[{{Stewart} {et~al.}(2011){Stewart}, {Kaufmann}, {Bullock}, {Barton},
  {Maller}, {Diemand}, \& {Wadsley}}]{stewart11}
{Stewart}, K.~R., {Kaufmann}, T., {Bullock}, J.~S., {et~al.} 2011, \apj, 738,
  39

\bibitem[{{Stocke} {et~al.}(2013){Stocke}, {Keeney}, {Danforth}, {Shull},
  {Froning}, {Green}, {Penton}, \& {Savage}}]{stocke13}
{Stocke}, J.~T., {Keeney}, B.~A., {Danforth}, C.~W., {et~al.} 2013, \apj, 763,
  148

\bibitem[{{Strawn} {et~al.}(2020){Strawn}, {Roca-F{\`a}brega}, {Mandelker},
  {Primack}, {Stern}, {Ceverino}, {Dekel}, {Wang}, \& {Dange}}]{Strawn20}
{Strawn}, C., {Roca-F{\`a}brega}, S., {Mandelker}, N., {et~al.} 2020, arXiv
  e-prints, arXiv:2008.11863

\bibitem[{{Sutherland} \& {Dopita}(1993)}]{sutherland93}
{Sutherland}, R.~S., \& {Dopita}, M.~A. 1993, \apjs, 88, 253

\bibitem[{{Tremonti} {et~al.}(2004){Tremonti}, {Heckman}, {Kauffmann},
  {Brinchmann}, {Charlot}, {White}, {Seibert}, {Peng}, {Schlegel}, {Uomoto},
  {Fukugita}, \& {Brinkmann}}]{Tremonti04}
{Tremonti}, C.~A., {Heckman}, T.~M., {Kauffmann}, G., {et~al.} 2004, \apj, 613,
  898

\bibitem[{Trujillo-Gomez {et~al.}(2015)Trujillo-Gomez, Klypin, Colin, Ceverino,
  Arraki, \& Primack}]{Trujillo15}
Trujillo-Gomez, S., Klypin, A., Colin, P., {et~al.} 2015, Monthly Notices of
  the Royal Astronomical Society, 446, 1140

\bibitem[{Tumlinson {et~al.}(2017)Tumlinson, Peeples, \&
  Werk}]{TumlinsonReview17}
Tumlinson, J., Peeples, M.~S., \& Werk, J.~K. 2017, Annual Review of Astronomy
  and Astrophysics, 55, 389

\bibitem[{{Tumlinson} {et~al.}(2011){Tumlinson}, {Thom}, {Werk}, {Prochaska},
  {Tripp}, {Weinberg}, {Peeples}, {O'Meara}, {Oppenheimer}, {Meiring}, {Katz},
  {Dav{\'e}}, {Ford}, \& {Sembach}}]{tumlinson11}
{Tumlinson}, J., {Thom}, C., {Werk}, J.~K., {et~al.} 2011, Science, 334, 948

\bibitem[{{van de Voort} \& {Schaye}(2012)}]{vandevoort+schaye12}
{van de Voort}, F., \& {Schaye}, J. 2012, \mnras, 423, 2991

\bibitem[{{van de Voort} {et~al.}(2019){van de Voort}, {Springel}, {Mandelker},
  {van den Bosch}, \& {Pakmor}}]{vandevoort19}
{van de Voort}, F., {Springel}, V., {Mandelker}, N., {van den Bosch}, F.~C., \&
  {Pakmor}, R. 2019, \mnras, 482, L85

\bibitem[{{van de Weygaert} \& {Bond}(2008)}]{VanBond08}
{van de Weygaert}, R., \& {Bond}, J.~R. 2008, {Clusters and the Theory of the
  Cosmic Web}, ed. M.~{Plionis}, O.~{L{\'o}pez-Cruz}, \& D.~{Hughes}, Vol. 740,
  335

\bibitem[{{Vander Vliet}(2017)}]{rachel_thesis}
{Vander Vliet}, J.~R. 2017, PhD thesis, New Mexico State University

\bibitem[{{Vergani} {et~al.}(2008){Vergani}, {Scodeggio}, {Pozzetti}, {Iovino},
  {Franzetti}, {Garilli}, {Zamorani}, {Maccagni}, {Lamareille}, {Le F{\`e}vre},
  {Charlot}, {Contini}, {Guzzo}, {Bottini}, {Le Brun}, {Picat}, {Scaramella},
  {Tresse}, {Vettolani}, {Zanichelli}, {Adami}, {Arnouts}, {Bardelli},
  {Bolzonella}, {Cappi}, {Ciliegi}, {Foucaud}, {Gavignaud}, {Ilbert},
  {McCracken}, {Marano}, {Marinoni}, {Mazure}, {Meneux}, {Merighi}, {Paltani},
  {Pell{\`o}}, {Pollo}, {Radovich}, {Zucca}, {Bondi}, {Bongiorno},
  {Brinchmann}, {Cucciati}, {de la Torre}, {Gregorini}, {Perez-Montero},
  {Mellier}, {Merluzzi}, \& {Temporin}}]{Vergani08}
{Vergani}, D., {Scodeggio}, M., {Pozzetti}, L., {et~al.} 2008, \aap, 487, 89

\bibitem[{{White} \& {Frenk}(1991)}]{white91}
{White}, S.~D.~M., \& {Frenk}, C.~S. 1991, \apj, 379, 52

\bibitem[{{White} \& {Rees}(1978)}]{white78}
{White}, S.~D.~M., \& {Rees}, M.~J. 1978, \mnras, 183, 341

\bibitem[{{Wiersma} {et~al.}(2009){Wiersma}, {Schaye}, \& {Smith}}]{wiersma09}
{Wiersma}, R. P.~C., {Schaye}, J., \& {Smith}, B.~D. 2009, \mnras, 393, 99

\bibitem[{{Zabl} {et~al.}(2020){Zabl}, {Bouch{\'e}}, {Schroetter}, {Wendt},
  {Contini}, {Schaye}, {Marino}, {Muzahid}, {Pezzulli}, {Verhamme}, \&
  {Wisotzki}}]{zabl20}
{Zabl}, J., {Bouch{\'e}}, N.~F., {Schroetter}, I., {et~al.} 2020, \mnras, 492,
  4576

\bibitem[{{Zahedy} {et~al.}(2019){Zahedy}, {Chen}, {Johnson}, {Pierce},
  {Rauch}, {Huang}, {Weiner}, \& {Gauthier}}]{Zahedy19}
{Zahedy}, F.~S., {Chen}, H.-W., {Johnson}, S.~D., {et~al.} 2019, \mnras, 484,
  2257

\bibitem[{Zolotov {et~al.}(2015)Zolotov, Dekel, Mandelker, Tweed, Inoue,
  DeGraf, Ceverino, Primack, Barro, \& Faber}]{zolotov15}
Zolotov, A., Dekel, A., Mandelker, N., {et~al.} 2015, Monthly Notices of the
  Royal Astronomical Society, 450, 2327

\end{thebibliography}

\end{document}